\documentclass[a4paper]{jpconf}
\usepackage{graphicx}
\begin{document}
\title{The influence of small-scale magnetic field
       on the evolution of inclination angle and precession damping
       in the framework of
       3-component model of neutron star
      }

\author{K Y Kraav$^1$, M V Vorontsov$^{1}$, 
        D P  Barsukov$^{2,1}$
       }

\address{$^1$ Peter the Great St. Petersburg Polytechnic University, Saint Petersburg, Russia } 
\address{$^2$ Ioffe Institute, Saint-Petersburg, Russia }

\ead{bars.astro@mail.ioffe.ru}

\begin{abstract}
The evolution of inclination angle and precession damping of radio pulsars is considered.
It is assumed that the neutron star consists of 
3 "freely"\  rotating components: the crust and two core components,
one of which contains pinned superfluid vortices.
We suppose that each component rotates as a rigid body.
Also the influence of the small-scale magnetic field on the star's braking
 process is examined.
Within the framework of this model the star simultaneously
can have glitch-like events combined with 
long-period precession (with periods $10-10^{4}$ years).
It is shown that the case of the small quantity of pinned superfluid vortices
seems to be more consistent with observations.
\end{abstract}

\section{Introduction}
Radio pulsars can be considered as exceptionally stable clocks.
But sometimes besides smooth braking their rotation suffers
some irregularities called glitches 
\cite{DAlessandro1996,Melatos2015}. 
Usually it is related to superfluid pinning. 
Pinned vortices do not allow superfluid 
to participate in pulsar braking.
It causes the sudden unpinning of vortices at some moments and 
the transfer of angular momentum from superfluid to the crust
\cite{Melatos2015}.
That is observed as a glitch.
There are some evidences that isolated neutron stars 
may precess with large periods $T_{p}$ of precession. 
Pulsar B1821-11 precesses with period 
$T_{p} \sim 500 \mbox{ days}$ \cite{Jones2016}.
Some other pulsars show periodic variations
which may be explained by precession with 
$T_{p} \sim 100 - 500 \mbox{ days}$
\cite{Johnston2015}. 
The changing of pulse profile of Crab pulsar 
also may be related to precession with $T_{p} \sim 10^{2} \mbox{ years}$
\cite{Arzamasskiy2015}.
The increase of radio luminosity of Geminga pulsar \cite{Malofeev2012}
may be related to precession with $T_{p} > 10 \mbox{ years}$ as well.
The non-regular pulse variations known as "red noise"\
\cite{DAlessandro1996,Melatos2015} 
with timescales $T \sim 1 \mbox{ month} - 10^{4} \mbox{ years}$
also can be related to precession
\cite{Biryukov2012}.
The problem is that the pinning of vortices leads to precession with periods 
$T_{p} \sim P \cdot (I_{tot} / L_{g} ) \sim (10^{2} - 10^{6}) \, P$
\cite{Link2006},
where $P$ is  pulsar period, $I_{tot}$ is moment of inertia of the star,
$L_{g}$ is the angular momentum of pinned superfluid.
Such precession seems to damped very quickly and it is incompatible with
existence of the long-period precession \cite{Link2006}.
In this paper we consider a model of rotating neutron star, proposed in \cite{Sedrakian1999}.
It allows the coexistence of long-term precession and quasi-glitch events.
We also take into account the influence of the small-scale magnetic
field on pulsar braking.

\section{Basic equations}
We assume that neutron star consists of 3 components which we will call the crust ($c$-component), $g$-component and $r$-component.
\\
{\bf The crust ($c$-component) }.
We suppose that it rotates as a rigid body with angular velocity
$\vec{\Omega}_{c}$.
It is the outer component so the angular velocity $\vec{\Omega}_{c}$ 
is the observed pulsar angular velocity $\vec{\Omega}$,
$\Omega = 2 \pi / P$.
We suppose that 
\begin{equation}
\vec{M}_{c} = I_{c} \vec{\Omega}_{c}
\mbox{\ and \ }
\dot{\vec{M}}_{c} = \vec{K}_{ext} + \vec{N}_{gc} + \vec{N}_{rc},
\label{Mc_def}
\end{equation}
where $\vec{M}_{c}$ is angular momentum of the crust, 
$I_{c}$ is its moment of inertia, 
$\vec{K}_{ext}$ is the external torque of magnetospheric origin
acting on the crust,
$\vec{N}_{gc}$ and $\vec{N}_{rc}$ are the torques 
acting on the crust due to its interaction with $g$ and $r$
components correspondingly.
\\
{\bf $g$-component}. It is the one of two inner components. 
We assume that it consists of normal matter 
rotating as a rigid body with angular velocity $\vec{\Omega}_{g}$
and superfluid matter 
firmly pinned to the normal matter
so that superfluid vortices rotate together with the normal matter
with angular velocity $\vec{\Omega}_{g}$:
\begin{equation}
\vec{M}_{g} = I_{g} \vec{\Omega}_{g} + \vec{L}_{g}
\mbox{, \ \ }
\dot{\vec{M}}_{g} = \vec{N}_{cg} + \vec{N}_{rg} 
\mbox{ \ and \ }
\dot{\vec{L}}_{g} = 
 \left[ \vec{\Omega}_{g} \times \vec{L}_{g} \right],
\label{Mg_def}
\end{equation}
where $\vec{M}_{g}$ is the total angular momentum of $g$-component,
$I_{g}$ is the moment of inertia of its normal matter,
$\vec{L}_{g}$ is angular momentum of pinned superfluid,
$\vec{N}_{cg}$ and $\vec{N}_{rg}$ are the torques 
acting on $g$-component due to its interaction with 
the crust and  $r$-component correspondingly. 
\\
{\bf $r$-component }. It is the second inner component.
We assume that it rotates as a rigid body
with angular velocity $\vec{\Omega}_{r}$:
\begin{equation}
\vec{M}_{r} = I_{r} \vec{\Omega}_{r}
\mbox{ \ and \ }
\dot{\vec{M}}_{r} = \vec{N}_{cr} + \vec{N}_{gr},
\label{Mr_def}
\end{equation}
where $\vec{M}_{r}$ is the angular momentum of $r$-component,
$I_{r}$ is its moment of inertia,
$\vec{N}_{cr}$ and $\vec{N}_{gr}$ are the torques 
acting on the $r$-component due to its interaction with 
the crust and $g$-component correspondingly.  
In the crust frame of references
the equations of rotation (\ref{Mc_def})-(\ref{Mr_def})
can be rewritten as
\begin{eqnarray}
\dot{\vec{\Omega}} & = & \vec{R}_{gc} + \vec{R}_{rc} + \vec{S}_{ext},
\label{dOmega_eqn}
\\
\dot{\vec{\mu}}_{cg} & + & [\vec{\Omega} \times \vec{\mu}_{cg} ]
+ [\vec{\Omega} \times \vec{\omega}_{g}]
+ [\vec{\mu}_{cg} \times \vec{\omega}_{g} ]
= \vec{R}_{cg} + \vec{R}_{rg} 
- \vec{R}_{gc} - \vec{R}_{rc} - \vec{S}_{ext},
\label{dmucg_eqn} 
\\
\dot{\vec{\mu}}_{cr} & + & [\vec{\Omega} \times \vec{\mu}_{cr} ]
= \vec{R}_{cr} + \vec{R}_{gr} - \vec{R}_{gc} - \vec{R}_{rc} - \vec{S}_{ext},
\label{dmucr_eqn}
\\
\dot{\vec{\omega}}_{g} & = &
\left[ \vec{\mu}_{cg} \times \vec{\omega}_{g} \right],
\label{dLg_eqn} 
\end{eqnarray}
where 
$\vec{\mu}_{ij} = \Omega_{j} - \Omega_{i}$, 
$\vec{N}_{ij} = - \vec{N}_{ji}$, 
$\vec{R}_{ij} = \vec{N}_{ij} / I_{j}$,
$i,j = c,g,r$,
$\vec{S}_{ext} = \vec{K}_{ext} / I_{c}$
and $\vec{\omega}_{g} = \vec{L}_{g} / I_{g}$.
\\
In the sake of simplicity we suppose that
\begin{equation}
\vec{N}_{ij} = - I_{j} \,
     \left(  \alpha_{ij} \mu^{||}_{ij} \vec{e}_{\Omega} 
           + \beta_{ij} \vec{\mu}^{\perp}_{ij}
           + \gamma_{ij} [ \vec{e}_{\Omega} \times \vec{\mu}^{\perp}_{ij} ]
     \right),
\label{Nij_def}
\end{equation}
where $\alpha_{ij},\beta_{ij},\gamma_{ij}$ are some constants,
$\vec{e}_{\Omega} = \vec{\Omega} / \Omega$ 
and we have introduced parallel  component $A^{||} = \vec{A} \cdot \vec{e}_{\Omega}$ and perpendicular component $\vec{A}^{\perp} = \vec{A} - A^{||} \vec{e}_{\Omega}$ for any vector $\vec{A}$.   

First let us consider the equilibrium state for zeroth
external torque $\vec{K}_{ext}=0$.
In this case,
the whole star rotates as a rigid body ($\vec{\mu}_{ij} = 0$)
and, hence, $\vec{R}_{ij} = 0$.
Equations (\ref{dOmega_eqn})-(\ref{dLg_eqn}) may be written as
\begin{equation}
\dot{\vec{\Omega}} = 0
\mbox{,\  }
\dot{\vec{\omega}}_{g} = 0 
\mbox{ and }
\vec{\omega}_{g} = \omega_{g} \vec{e}_{\Omega}.
\end{equation}
Let us further consider a small perturbation to the equilibrium state.
We will treat values $\vec{\mu}_{ij}$, $\vec{\omega}_{g}^{\perp}$
and $\vec{S}_{ext}$
as small perturbations and neglect any term quadratic in these values. 
Hence, equations (\ref{dOmega_eqn})-(\ref{dLg_eqn}) may be written as
\begin{eqnarray}
\dot{\Omega} & = & R_{gc}^{||} + R_{rc}^{||} + S_{ext}^{||},
\label{dOmega_linear_pp_eqn}
\\
\dot{\mu}_{cg}^{||} & = &
  R_{cg}^{||} + R_{rg}^{||}
- R_{gc}^{||} - R_{rc}^{||} - S_{ext}^{||},
\label{dmucg_linear_pp_eqn} 
\\
\dot{\mu}_{cr}^{||} & = &
R_{cr}^{||} + R_{gr}^{||} - R_{gc}^{||} - R_{rc}^{||} - S_{ext}^{||},
\label{dmucr_linear_pp_eqn}
\\
\dot{\omega}_{g}^{||} & = & 0,
\label{dLg_linear_pp_eqn}  
\\
\Omega \dot{\vec{e}}_{\Omega} & = & 
\vec{R}_{gc}^{\perp} + \vec{R}_{rc}^{\perp} + \vec{S}_{ext}^{\perp},
\label{dOmega_linear_perp_eqn}
\\
\dot{\vec{\mu}}_{cg}^{\perp} & - & 
  (\omega_{g}^{||} - \Omega) \, 
  [ \vec{e}_{\Omega} \times \vec{\mu}_{cg}^{\perp} ]
+ [\vec{\Omega} \times \vec{\omega}_{g}^{\perp}]
= \vec{R}_{cg}^{\perp} + \vec{R}_{rg}^{\perp} 
- \vec{R}_{gc}^{\perp} - \vec{R}_{rc}^{\perp} - \vec{S}_{ext}^{\perp},
\label{dmucg_linear_perp_eqn} 
\\
\dot{\vec{\mu}}_{cr}^{\perp} & + & 
[\vec{\Omega} \times \vec{\mu}_{cr}^{\perp} ]
= \vec{R}_{cr}^{\perp} + \vec{R}_{gr}^{\perp} 
- \vec{R}_{rc}^{\perp} - \vec{R}_{gc}^{\perp} - \vec{S}_{ext}^{\perp},
\label{dmucr_linear_perp_eqn}
\\
\dot{\vec{\omega}}_{g}^{\perp} & = & 
- \frac{\omega_{g}^{||}}{\Omega} \,
  \left(
     \vec{R}_{gc}^{\perp} + \vec{R}_{rc}^{\perp} + \vec{S}_{ext}^{\perp} 
   + [ \vec{\Omega} \times \vec{\mu}_{cg} ]
  \right).
\label{dLg_linear_perp_eqn}   
\end{eqnarray}

\begin{figure}[h]
\begin{minipage}{14pc}
\includegraphics[width=10pc]{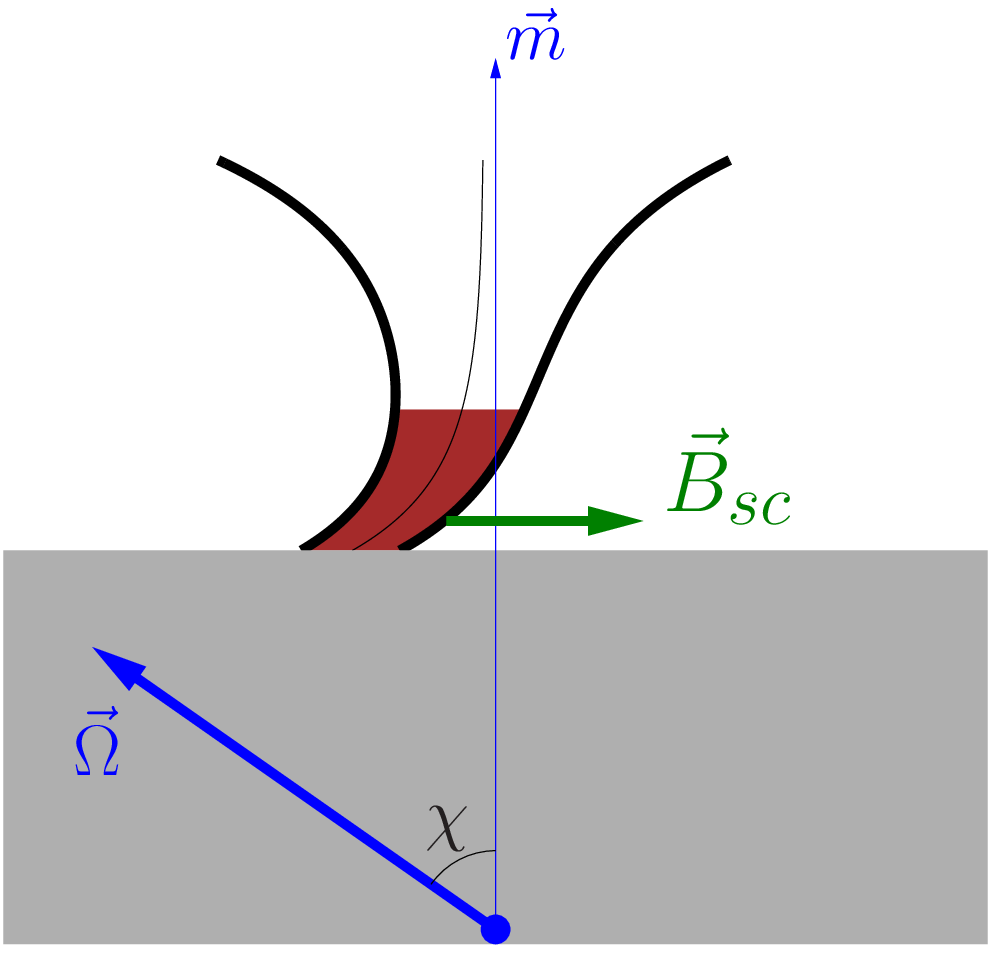}
\caption{\label{fig_Bsc}
          The direction of small-scale 
          magnetic field $\vec{B}_{sc}$ and 
          the direction of dipolar magnetic moment $\vec{m}$.
          "Inner gap"\  is shown by brown area, neutron star is
          shown by grey area.
        }
\end{minipage}
\hspace{2pc}%
\begin{minipage}{14pc}
\includegraphics[width=10pc,angle=270]{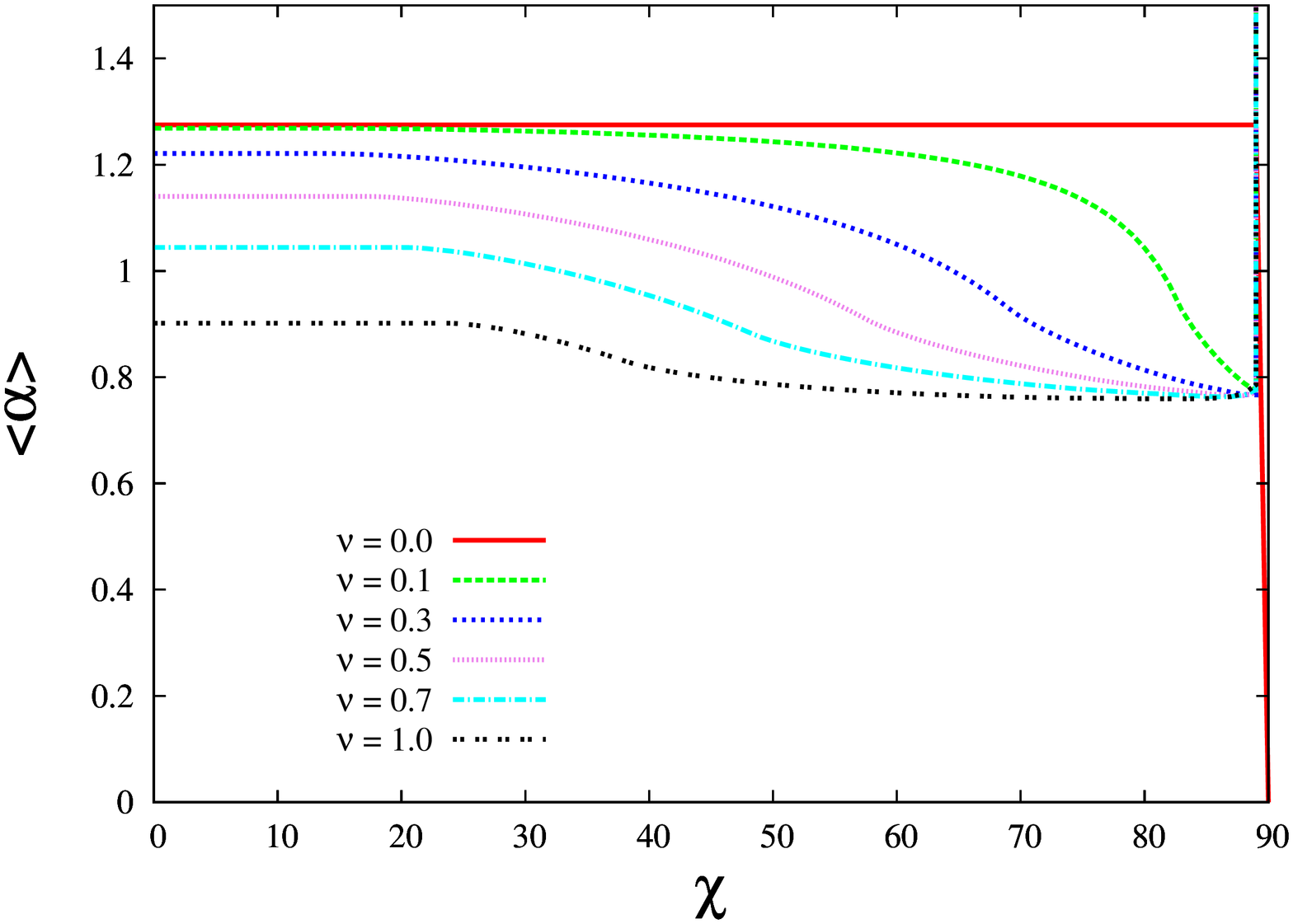}
\caption{\label{fig_av_alpha}
          The dependence of parameter $<\alpha>$ on inclination angle $\chi$
          taken from \cite{Polyakova2009}
        }
\end{minipage} 
\end{figure}

\begin{figure}[h]
\begin{minipage}{14pc}
\includegraphics[width=10pc]{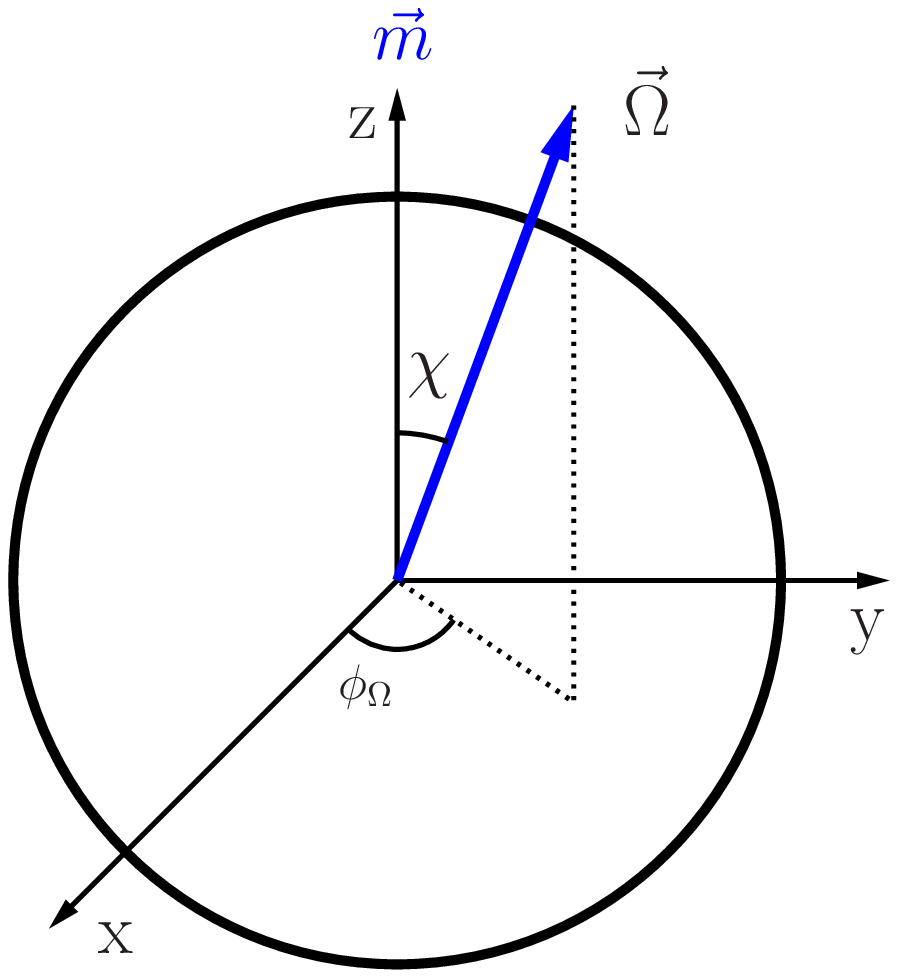}
\caption{\label{fig_angles_def}
         The definitions of angles $\chi$ and $\phi_{\Omega}$.
        } 
\end{minipage}
\hspace{2pc}%
\begin{minipage}{14pc}
\includegraphics[width=10pc]{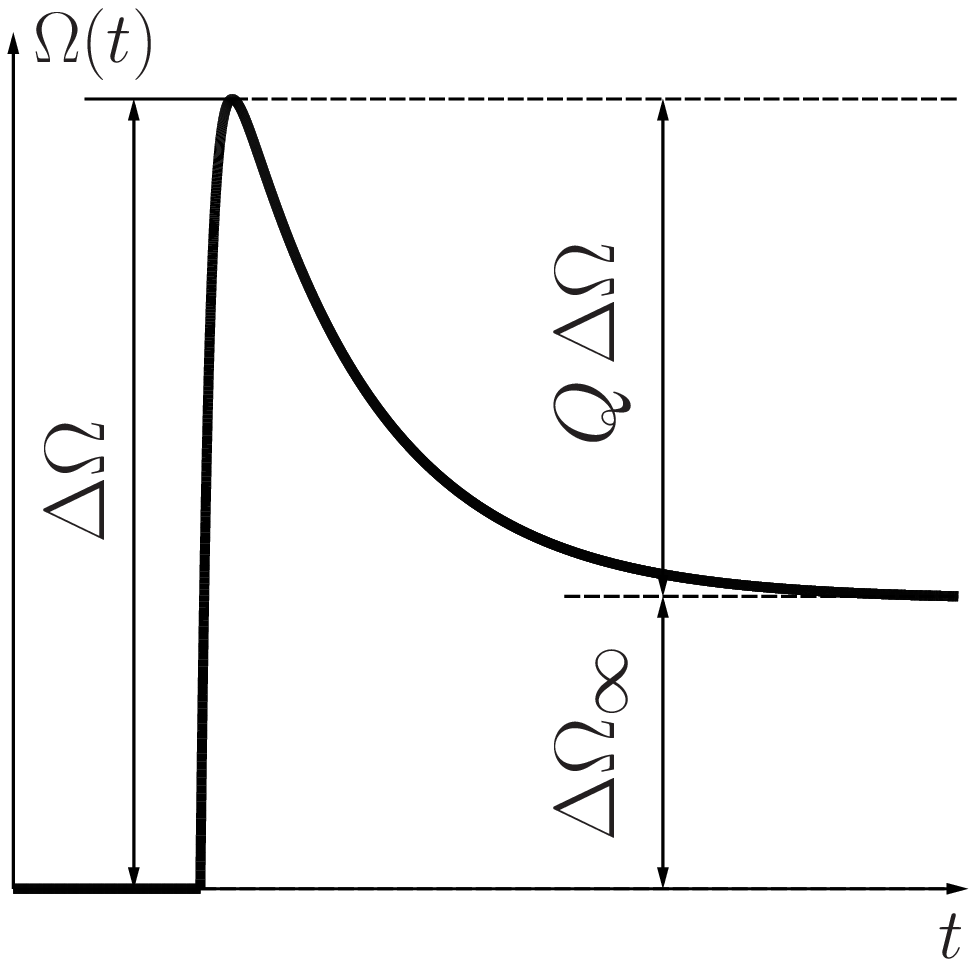}
\caption{\label{fig_glitch}
          The evolution of angular velocity $\Omega(t)$
          during a quasi-glitch event.
          The pulsar braking is neglected. 
          %The evolution of departure $\Omega_{(h)}$ of angular velocity 
          %from its quasistatic value during quasi-glitch event.
        } 
\end{minipage} 
\end{figure}

In order to calculate the magnetospheric torque $\vec{K}_{ext}$
acting on the crust we use the model proposed in
\cite{Polyakova2009}.
It is assumed that neutron star is braking simultaneously
by both magnitodipolar and current losses. Hence,
\begin{equation}
\vec{K}_{ext} = - \frac{\tilde{I}_{tot}}{\tau_{0}} \,
  \left( \vec{e}_{\Omega} - (1-\alpha) \cos\chi \, \vec{e}_{m}
       - R_{eff} [\vec{e}_{\Omega} \times \vec{e}_{m}] 
  \right),
\label{Kext_def}
\end{equation}
where $\vec{m} = m \vec{e}_{m}$ is the dipolar magnetic moment
of neutron star,
$\chi$ is the inclination angle 
(the angle between $\vec{e}_{\Omega}$ and $\vec{e}_{m}$, 
see fig.\ref{fig_angles_def}),
$\tau_{0} = \frac{3}{2} \frac{c^{3}}{m^{2} \Omega^{3}}
\tilde{I}_{tot}$,
$\tilde{I}_{tot} = I_{c} + I_{g} + I_{r}$,
the coefficient 
$R_{eff}$ is related to the magnetic field inertia
\cite{Zheltoukhov2014,Goglichidze2015}.
In the paper we assume that
$R_{eff} = \frac{9}{10} \frac{c}{\Omega r_{ns}}
\sim 5 \cdot 10^{3} \left( \frac{P}{1s} \right)
$ 
\cite{Melatos2000},
where $r_{ns}$ is neutron star radius.
The coefficient $\alpha$ is related to the 
value of the current flowing through the pulsar tubes.
In this paper we assume that 
there are only "inner gaps"\ 
with free electron emission from neutron star surface
in pulsar tube.
Hence, the magnitude of the current depends on the structure of surface
small-scale magnetic field (see fig. \ref{fig_Bsc}).
The value of $\alpha$ averaged over precession angle $\phi_{\Omega}$ (see fig \ref{fig_angles_def}) 
\begin{equation}
 <\alpha>(\chi) = \frac{1}{2\pi} 
  \int_{0}^{2\pi} \alpha(\chi,\phi_{\Omega}) d\phi_{\Omega}
\end{equation}
is shown in fig \ref{fig_av_alpha}.
Here, $\nu = B_{sc} / B_{dip}$, 
$B_{sc}$ is the induction of small-scale magnetic field,
$B_{dip} = 2 m / r_{ns}^{3}$ is the induction of
dipolar field on the magnetic pole of neutron star.


\section{Quasistatic approximation}
Firstly let us take into account that
$\tau_{rel} \ll \tau_{0}$,
where 
$\tau_{rel} \sim 
{\mathrm{max}}\left( 1 /\alpha_{ij}, 1/\beta_{ij}, 1/\gamma_{ij} \right)
\sim (1 - 10^{7}) s$ is relaxation time
\cite{Melatos2015}.
Hence, we can consider the rotation of neutron star
under the small slowly varying torque $\vec{K}_{ext}$.
In this case, it is possible to neglect terms
$\dot{\vec{\mu}}_{ij}$ and $\dot{\vec{\omega}}_{g}^{\perp}$ 
in (\ref{dOmega_linear_pp_eqn})-(\ref{dLg_linear_perp_eqn}). 
Using (\ref{dOmega_linear_pp_eqn})-(\ref{dmucr_linear_pp_eqn})
we can immediately obtain that
\begin{equation}
\dot{\Omega} = \frac{ K_{ext}^{||} }{ \tilde{I}_{tot} }
\end{equation}
so the star brakes as a rigid body with 
moment of inertia equal to $\tilde{I}_{tot}$.
Equations (\ref{dmucg_linear_pp_eqn}) and (\ref{dmucr_linear_pp_eqn}) give us
\begin{eqnarray}
\mu_{cg}^{||} & = & - \frac{ S_{ext}^{||} I_{c} }{ \tilde{I}_{tot} } \cdot
  \frac{ \alpha_{cr} + \alpha_{rg} + \alpha_{gr} }
       { \alpha_{cg} \alpha_{cr} + \alpha_{rg} \alpha_{cr} 
         + \alpha_{cg} \alpha_{gr} 
       },
\label{mucg_pp_qs} 
\\
\mu_{cr}^{||} & = & - \frac{ S_{ext}^{||} I_{c} }{ \tilde{I}_{tot} } \cdot
  \frac{ \alpha_{cg} + \alpha_{rg} + \alpha_{gr} }
       { \alpha_{cg} \alpha_{cr} + \alpha_{rg} \alpha_{cr} 
         + \alpha_{cg} \alpha_{gr} 
       }.
\label{mucr_pp_qs} 
\end{eqnarray}
Let us  introduce a complex number $A^{\perp}$ associated to arbitrary perpendicular vector $\vec{A}^{\perp}$ in the following way:
if $\vec{e}_{z} = \vec{e}_{\Omega}$ then
$\vec{A}^{\perp} = A_{x} \vec{e}_{x} + A_{y} \vec{e}_{y}$
and $A^{\perp} = A_{x} + i A_{y}$.
Hence, equations (\ref{dmucg_linear_perp_eqn})-(\ref{dLg_linear_perp_eqn})
give us
\begin{eqnarray}
\mu_{cg}^{\perp} & = & \frac{S_{ext}^{\perp}}{\Delta_{\perp}} \cdot
               \left( i\Omega + \xi_{cr} + \xi_{gr} \right),
\label{mucg_perp_qs}
\\
\mu_{cr}^{\perp} & = & \frac{S_{ext}^{\perp}}{\Delta_{\perp}} \cdot
               \left( i\Omega + \xi_{gr} \right),
\label{mucr_perp_qs}
\\
\omega_{g}^{\perp} & = & 
  \frac{S_{ext}^{\perp}}{\Omega \, \Delta_{\perp}} \cdot
  \left( \omega_{g}^{||} \, \left( i\Omega + \xi_{cr} + \xi_{gr} \right)
       - \Omega \, \xi_{cg}
       + \xi_{cg} \xi_{cr} + \xi_{cg} \xi_{gr} + \xi_{rg} \xi_{cr}
  \right),
\label{Lg_perp_qs}
\end{eqnarray}
where $\xi_{pq} = \beta_{pq} + i \gamma_{pq}$ and  
\begin{equation}
\Delta_{\perp} = \Omega^{2}
 - i \Omega \, ( \xi_{cr} + \xi_{rc} + \xi_{gr} + \xi_{gc} )
 - \left( \xi_{gc} \xi_{cr} + \xi_{gc} \xi_{gr} + \xi_{rc} \xi_{gr}
   \right).
\end{equation}
In the case of weak viscosity
$|\xi_{pq}| \ll \Omega$ we have 
\begin{equation}
\vec{\mu}_{cg}^{\perp} \approx \frac{1}{\Omega} \,
       [ \vec{e}_{\Omega} \times \vec{S}_{ext}^{\perp} ]
\mbox{, \ }
\vec{\mu}_{cr}^{\perp} \approx \frac{1}{\Omega} \,
       [ \vec{e}_{\Omega} \times \vec{S}_{ext}^{\perp} ] 
\mbox{\ and \ } 
\vec{\omega}_{g}^{\perp} \approx
       \frac{\omega_{g}^{||}}{\Omega^{2}} \,
       [ \vec{e}_{\Omega} \times \vec{S}_{ext}^{\perp} ]. 
\end{equation}
In the last equation we also assume that $|\xi_{pq}| \ll \omega_{g}^{||}$.
\\
Taking into account equations (\ref{mucg_pp_qs}) - (\ref{mucr_perp_qs})
and (\ref{Kext_def}) one can rewrite
equations (\ref{dOmega_linear_pp_eqn}) and (\ref{dOmega_linear_perp_eqn})
 as
\begin{eqnarray}
\dot{\Omega} & = &  - \frac{\Omega}{\tau_{0}}
           \left( \sin^{2}\chi + \alpha \cos^{2}\chi \right),
\label{PB_dO_dt_eqn}
\\
\dot{\chi} & = &
 - \frac{\tilde{I}_{tot}}{I_{c}} \, \frac{1}{\tau_{0}} \,
   \sin\chi \, \cos\chi \cdot
   \left( B \, (1-\alpha) - \Gamma \, R_{eff} \right),
\label{PB_dchi_dt_eqn}
\\
\dot{\phi}_{\Omega} & = &
 - \frac{\tilde{I}_{tot}}{I_{c}} \, \frac{1}{\tau_{0}} \,
   \cos\chi \cdot
   \left( B \, R_{eff} + \Gamma \, (1-\alpha) \right),
\label{PB_dphi_dt_eqn}
\end{eqnarray} 
where coefficients $B$ and $\Gamma$ are defined as
\begin{equation}
B + i\Gamma = \frac{\Omega}{\Delta_{\perp}} \cdot
         \left(\Omega - i(\xi_{cr} + \xi_{gr}) \right).
\label{BG_def}
\end{equation}
In the case of weak viscosity
$|\xi_{pq}| \ll \Omega$ we have
\begin{equation}
B \approx 1 - \frac{ \gamma_{gc} + \gamma_{rc} }{ \Omega }
\mbox{ \  and \  }
\Gamma \approx \frac{ \beta_{gc} + \beta_{rc} }{ \Omega }.
\end{equation}
And consequently equation (\ref{PB_dphi_dt_eqn}) gives that
as long as inclination angle $\chi$ differs 
from $0^{\circ}$ and $90^{\circ}$
the star will precess
with period
\begin{equation}
T_{p} = \frac{ \tau_{0} }{ R_{eff} } \cdot
        \frac{ I_{c} }{ \tilde{I}_{tot} } \cdot
        \frac{ 2\pi }{ \cos\chi }
  \sim (10^{-5} - 10^{-3}) \, \tau_{0}
  \sim 10 - 10^{3} \mbox{ years}.
\end{equation}
Also it is worth to note that as long as
quasistatic approximation is valid
the pinned superfluid $\vec{L}_{g}$ does not influence
on star rotation.
\\ 
We suppose that the period of precession
$T_{p} \ll \left( I_{c} /\tilde{I}_{tot} \right) \cdot \tau_{0} \ll \tau_{0}$.
Hence,
we can average equations (\ref{PB_dO_dt_eqn}) and (\ref{PB_dchi_dt_eqn})
over precession
obtaining equation
\begin{equation}
\frac{d\chi}{d P} = 
 - \frac{1}{P} \cdot \frac{\tilde{I}_{tot}}{I_{c}} \cdot
   \sin\chi \cos\chi \cdot
   \frac{ B \, (1-< \alpha >) - \Gamma \, R_{eff} }
        { \sin^{2}\chi + < \alpha > \cos^{2}\chi }.
\label{PB_dchi_dP_base_eqn} 
\end{equation}

\section{Quasi-glitch events}
According to equation (\ref{Lg_perp_qs}) the direction of $\vec{L}_{g}$
follows  the vector $\vec{e}_{\Omega}$
so the pinned superfluid vortices are directed almost
along the crust rotation axis.
Consequently, the direction of pinned vortices precesses together with $\vec{e}_{\Omega}$.
However, due to ideal pinning (see the last equation in (\ref{Mg_def})) 
the magnitude of vector $\vec{L}_{g}$ does not change at all.
Hence, due to pulsar braking the difference between
velocities of superfluid and normal matter will grow.
So the glitch must occur in $g$-component at some moment.
We suppose that the glitch occurs and relaxes at time-scale 
much smaller than the period of precession $T_{p}$.
It allows us to consider $\vec{e}_{\Omega}$ as a constant vector 
in equations (\ref{dmucg_linear_pp_eqn}), (\ref{dmucr_linear_pp_eqn}), (\ref{dmucg_linear_perp_eqn})-(\ref{dLg_linear_perp_eqn}) 
and, for simplicity, neglect the acting of external torque $\vec{S}_{ext}$.

Let us assume that before glitch the neutron star rotates as a rigid body.
At some moment ($t=0$) a small amount of angular momentum $\Delta \vec{L}_{g} = \Delta L_{g} \vec{e}_{\Omega}$ is is transfered from superfluid part to normal part of $g$-component.
Then the solution of equations
(\ref{dOmega_linear_pp_eqn})-(\ref{dmucr_linear_pp_eqn})
gives us (see fig. \ref{fig_glitch})
\begin{equation}
\Omega(t) = \Delta\Omega
\left( 1 - e^{-p_{+}t} - Q (1-e^{-p_{-}t}) \right),
\end{equation}
where 
$\Delta\Omega = \frac{\Delta \Omega_{\infty}}{1-Q}$,
$\Delta\Omega_{\infty} = \frac{ \Delta L_{g}}{\tilde{I}_{tot}}$,
the coefficients $p_{+}$ and $p_{-}$ ($p_{+} > p_{-}$)
are the roots of equation
\begin{eqnarray}
p^{2} & - &
\left( \alpha_{cg} + \alpha_{rg} + \alpha_{cr} + \alpha_{gr} 
     + \alpha_{gc} + \alpha_{rc}
\right) \, p 
+
\nonumber
\\
 & + &
  \left( \alpha_{gc} + \alpha_{rg} + \alpha_{cg} \right) \cdot
  \left( \alpha_{cr} + \alpha_{gr} + \alpha_{rc} \right)
+ \left( \alpha_{rc} - \alpha_{rg} \right) \cdot
  \left( \alpha_{gr} - \alpha_{gc} \right)
= 0
\label{p_def} 
\end{eqnarray}
and 
\begin{equation}
Q = \frac{ \tilde{I}_{tot} \alpha_{cg} - I_{c} p_{+} }
         { \tilde{I}_{tot} \alpha_{cg} - I_{c} p_{-} }
\end{equation}
In the case of 
$\alpha_{cg} \gg 
 \left( 1 + \frac{I_{c}}{I_{r}} \right) \alpha_{rc},
 \left( 1 + \frac{I_{g}}{I_{r}} \right) \alpha_{rg}
$
we have
\begin{equation}
p_{+} \approx \left( 1 + \frac{I_{g}}{I_{c}} \right) \alpha_{cg}
\mbox{, \ }
p_{-} \approx \frac{\tilde{I}_{tot}}{I_{c} + I_{g}} \,
               \left( \alpha_{cr} + \alpha_{gr} \right)
\mbox{\  and \ }
1 - Q \approx \frac{I_{c}+I_{g}}{\tilde{I}_{tot}}
\end{equation}
If one want to relate these quasi-glitch event to
observed glitches then
$1 / p_{+}$ and $1 / p_{-}$ should be interpreted as the glitch growth and relaxation times respectively.
We obtain that
$1 / p_{+} \leq 1 \mbox{ min}$ \cite{Dodson2002,Melatos2015} and
$1 / p_{-} \sim 1 - 10^{2} \mbox{ days}$ \cite{DAlessandro1996}.
Unfortunately, $1-Q \sim 10^{-2}-10^{-1}$ in our model.
That may be not so bad for glitches in some pulsars
like Crab ($Q \geq 0.8$ \cite{Melatos2015})
or J0205+6446 ($Q \approx 0.77$ \cite{Gurgercinoglu2017}), 
but obviously contradicts glitches in most pulsars for which $Q \ll 1$ \cite{Gurgercinoglu2017,Yu2013}.
In particular, our model does not describe glitches in Vela pulsar 
$Q \leq 0.2$ \cite{Melatos2015}.

\section{Results}
\begin{figure}[h]
\parbox{0.45\hsize}{
 \includegraphics[angle=270,width=0.9\hsize]{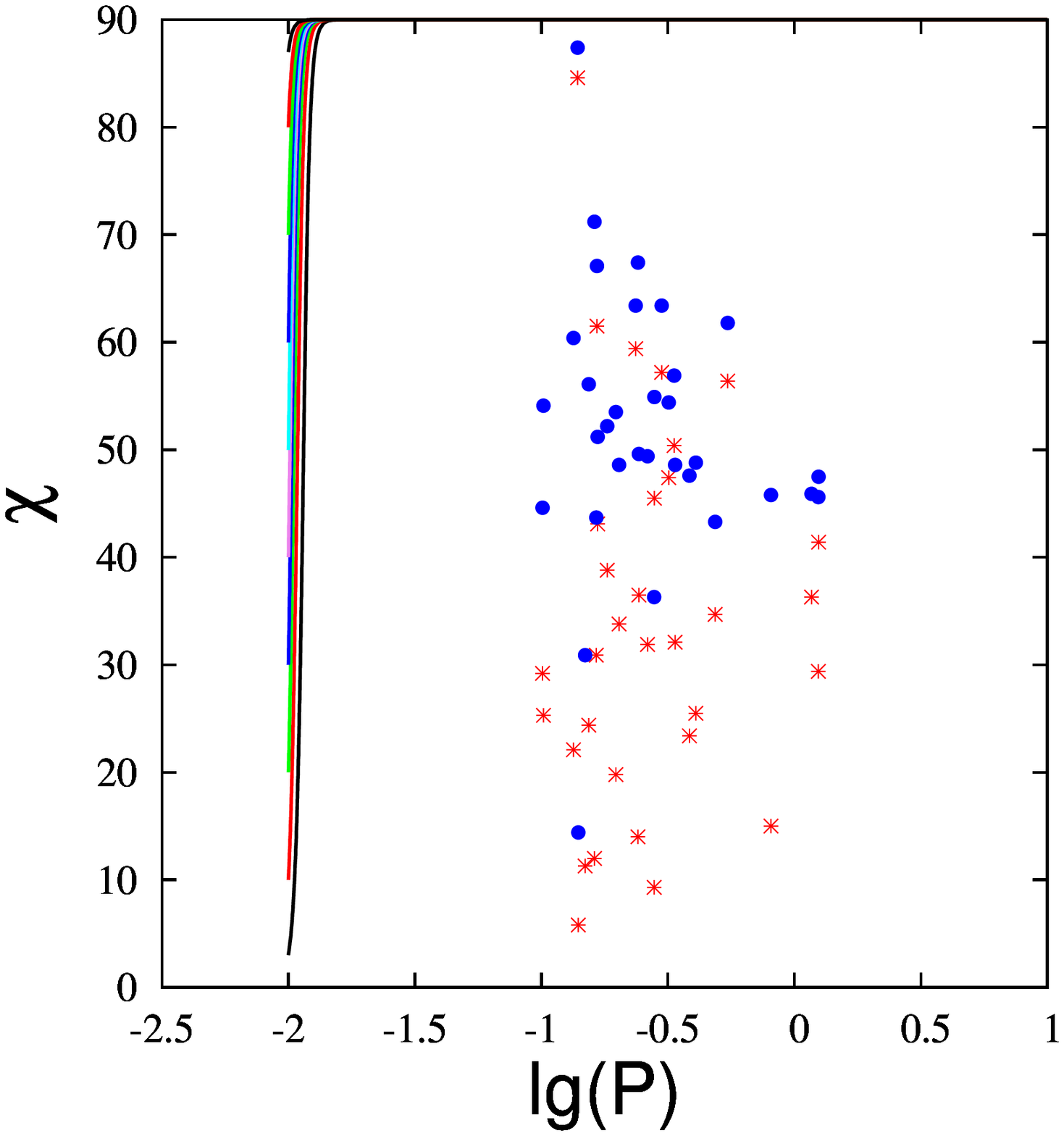} 
} % end parbox
\parbox{0.45\hsize}{
 \includegraphics[angle=270,width=0.9\hsize]{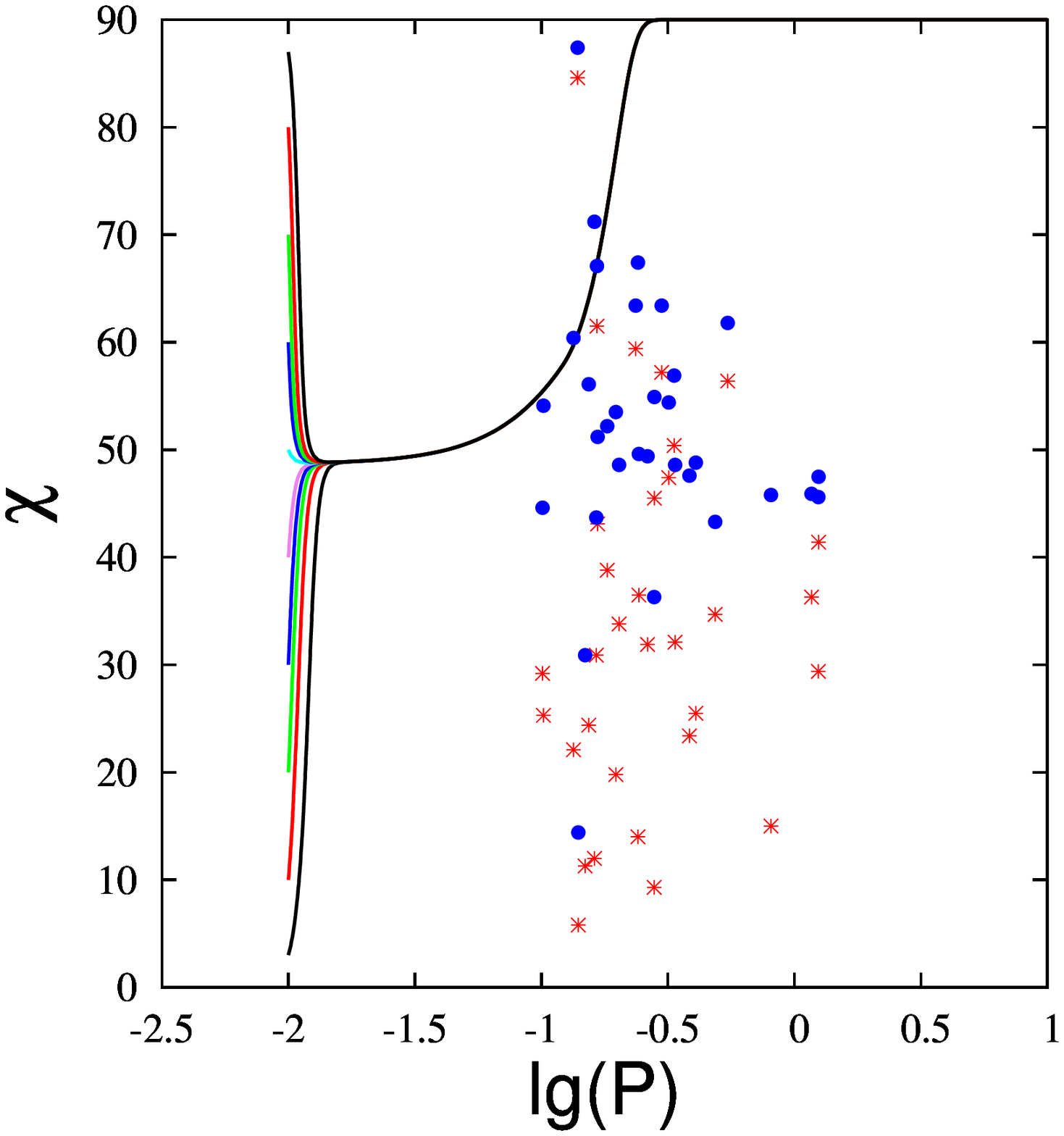}  
} % end parbox   
\caption{
   The evolution of inclination angle $\chi$
   for $\nu=0.0$ (left panel) and $\nu=0.5$ (right panel),
   $I_{c} / \tilde{I}_{tot} = 10^{-2}$,
   $I_{g} / \tilde{I}_{tot} = 10^{-3}$,
   $\alpha_{cg} = 10^{-1} \mbox{ s}^{-1}$, $\sigma = 10^{-10}$.
   Observed inclination angles $\beta_{2}$ at 10 cm 
   is taken from \cite{Nikitina2011_1},
   red stars correspond to $C>0$, blue dots corresponds to $C<0$.  
}
\label{fig_Icm2_init_1}
\end{figure}

\begin{figure}[h]
\parbox{0.45\hsize}{
 \includegraphics[angle=270,width=0.9\hsize]{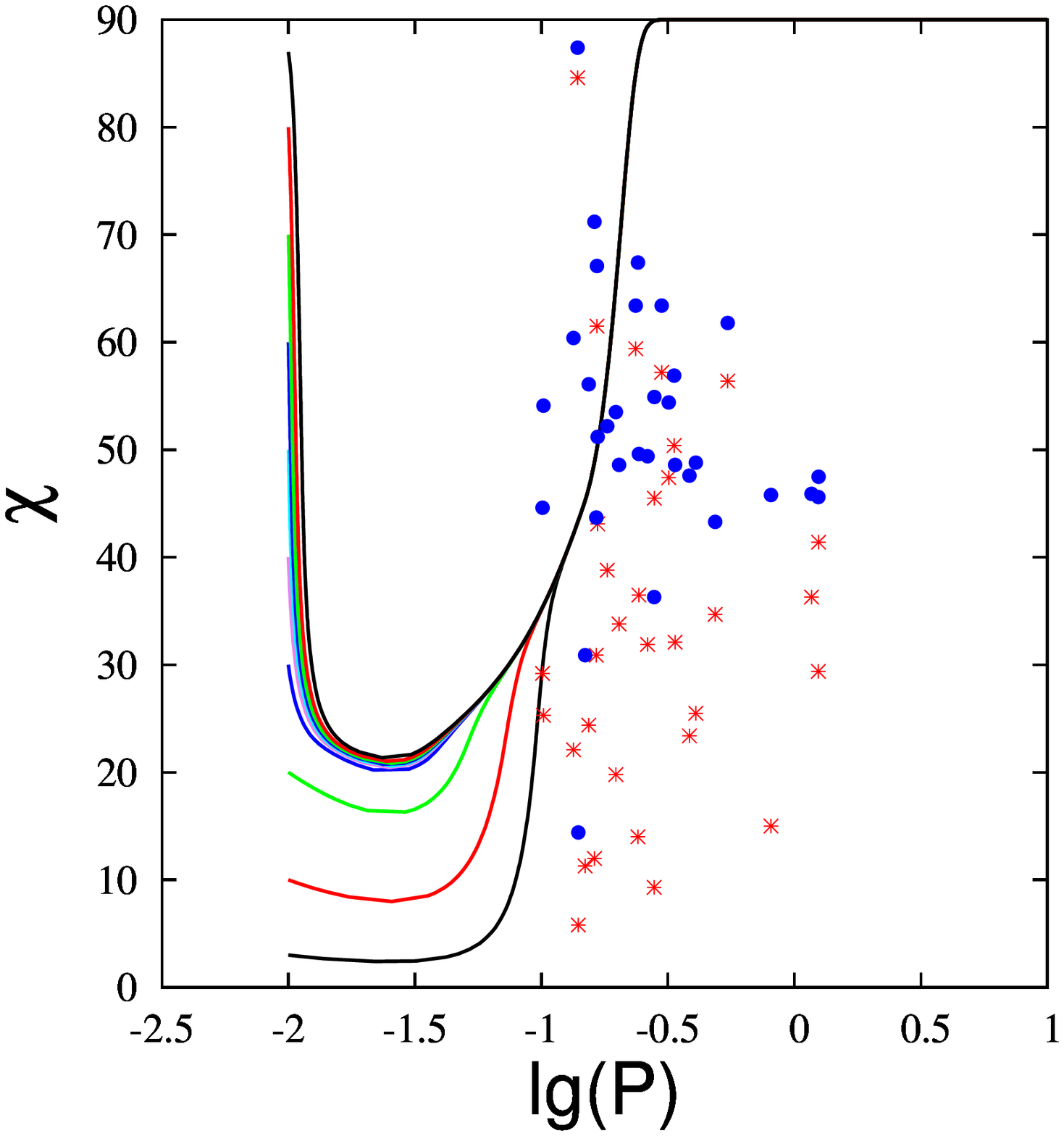} 
} % end parbox
\parbox{0.45\hsize}{
 \includegraphics[angle=270,width=0.9\hsize]{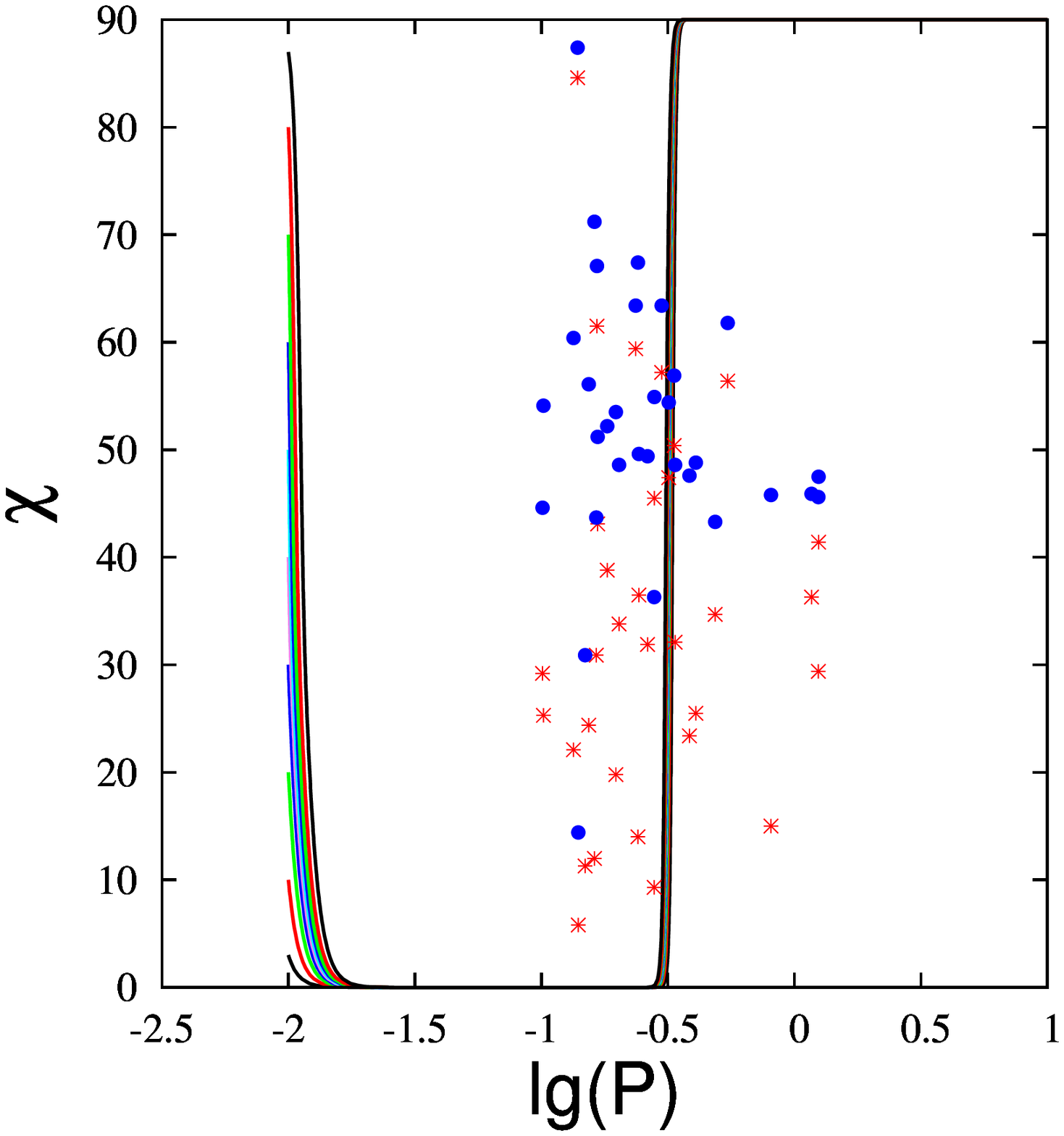}  
} % end parbox   
\caption{
   The same as in fig. \ref{fig_Icm2_init_1} but
   for $\nu = 0.8$ (left panel) and
   $\nu = 1.0$ (right panel).
}
\label{fig_Icm2_init_2}
\end{figure}

\begin{figure}[h]
\parbox{0.45\hsize}{
 \includegraphics[angle=270,width=0.9\hsize]{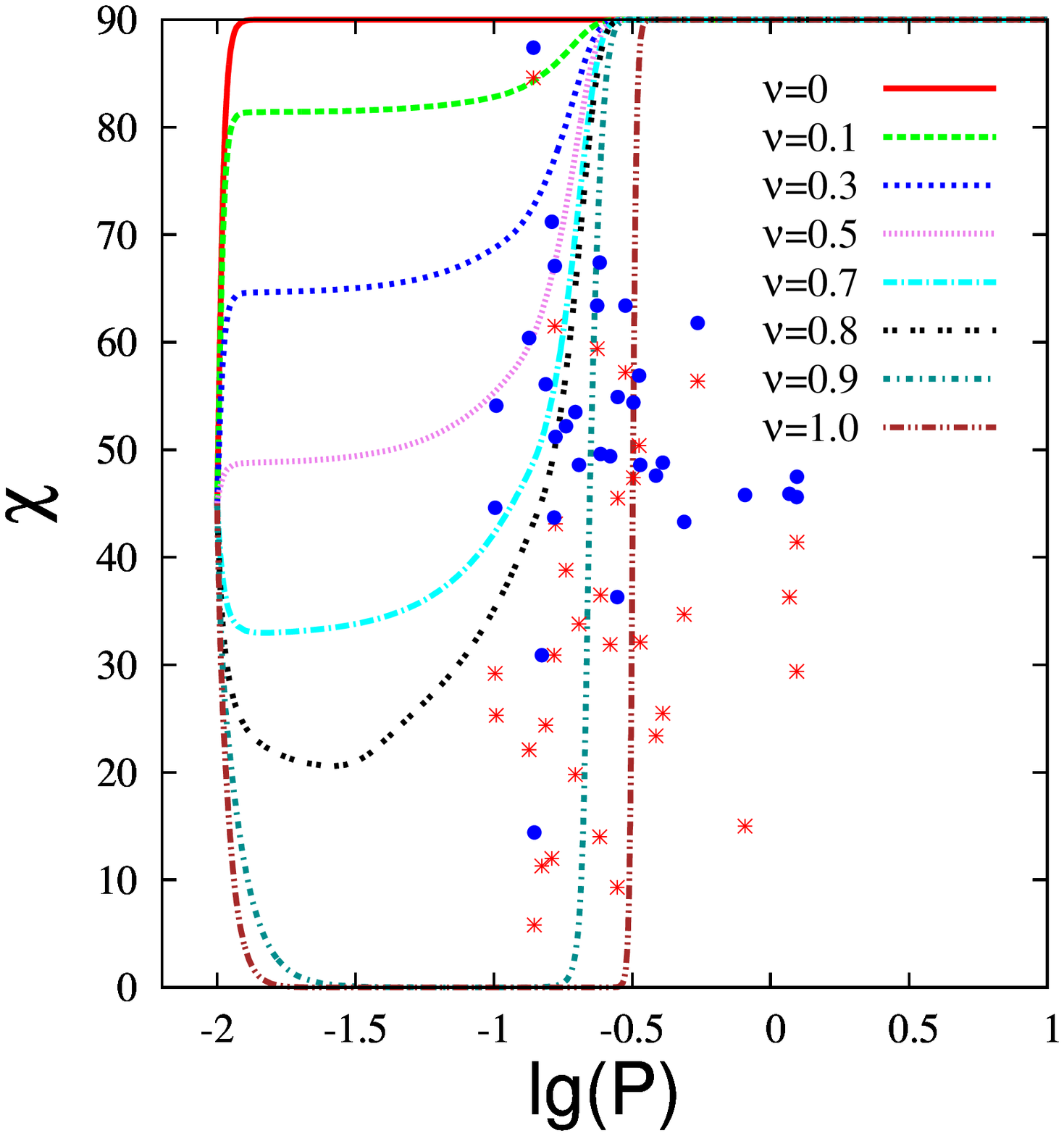} 
} % end parbox
\parbox{0.45\hsize}{
 \includegraphics[angle=270,width=0.9\hsize]{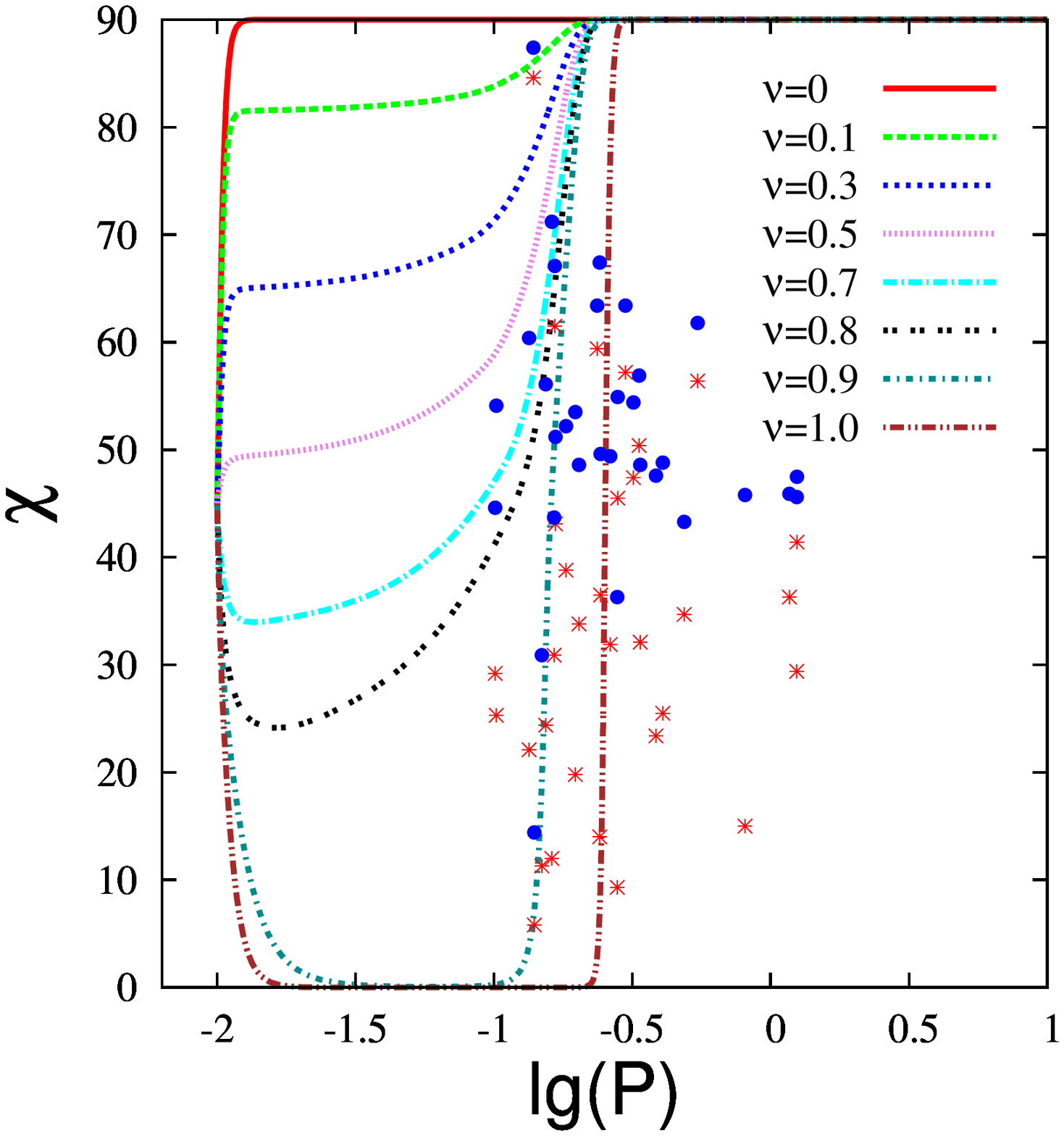}  
} % end parbox   
\caption{
   The evolution of inclination angle $\chi$ for different
   $\nu$, 
   $\sigma = 10^{-10}$ (left panel), 
   $\sigma = 10^{-6}$ (right panel),
   $\alpha_{cg} = 10^{-1} \mbox{ s}^{-1}$,
   $I_{c} / \tilde{I}_{tot} = 10^{-2}$,
   $I_{g} / \tilde{I}_{tot} = 10^{-3}$.
   Initial inclination angle $\chi = 45^{\circ}$, initial period $P = 10 \mbox{ ms}$.      
   Observed inclination angles $\beta_{2}$ at 10 cm 
   is taken from \cite{Nikitina2011_1},
   red stars correspond to $C>0$, blue dots to $C<0$. 
}
\label{fig_Icm2_nu}
\end{figure}

\begin{figure}[h]
\parbox{0.45\hsize}{
 \includegraphics[angle=270,width=0.9\hsize]{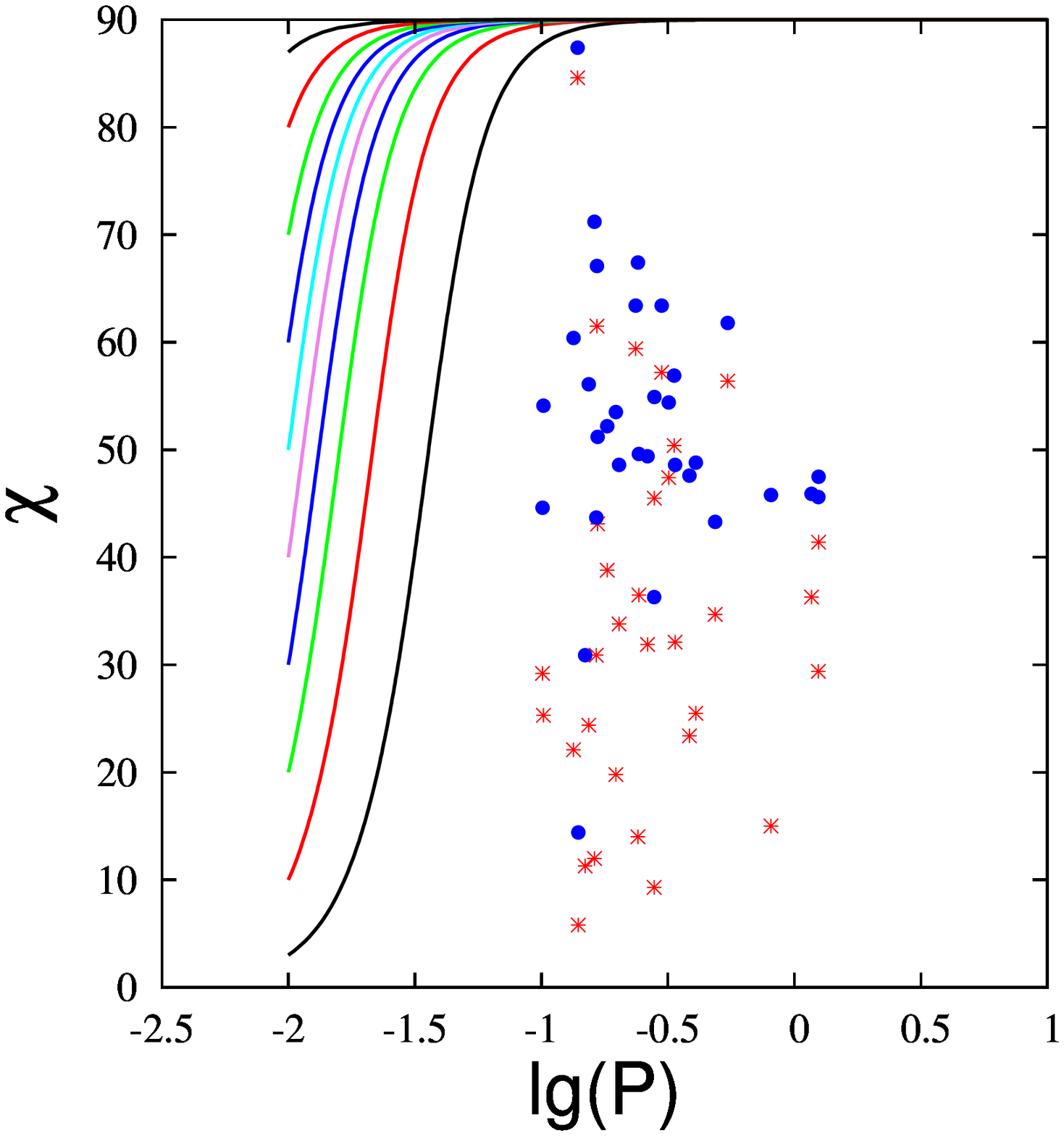} 
} % end parbox
\parbox{0.45\hsize}{
 \includegraphics[angle=270,width=0.9\hsize]{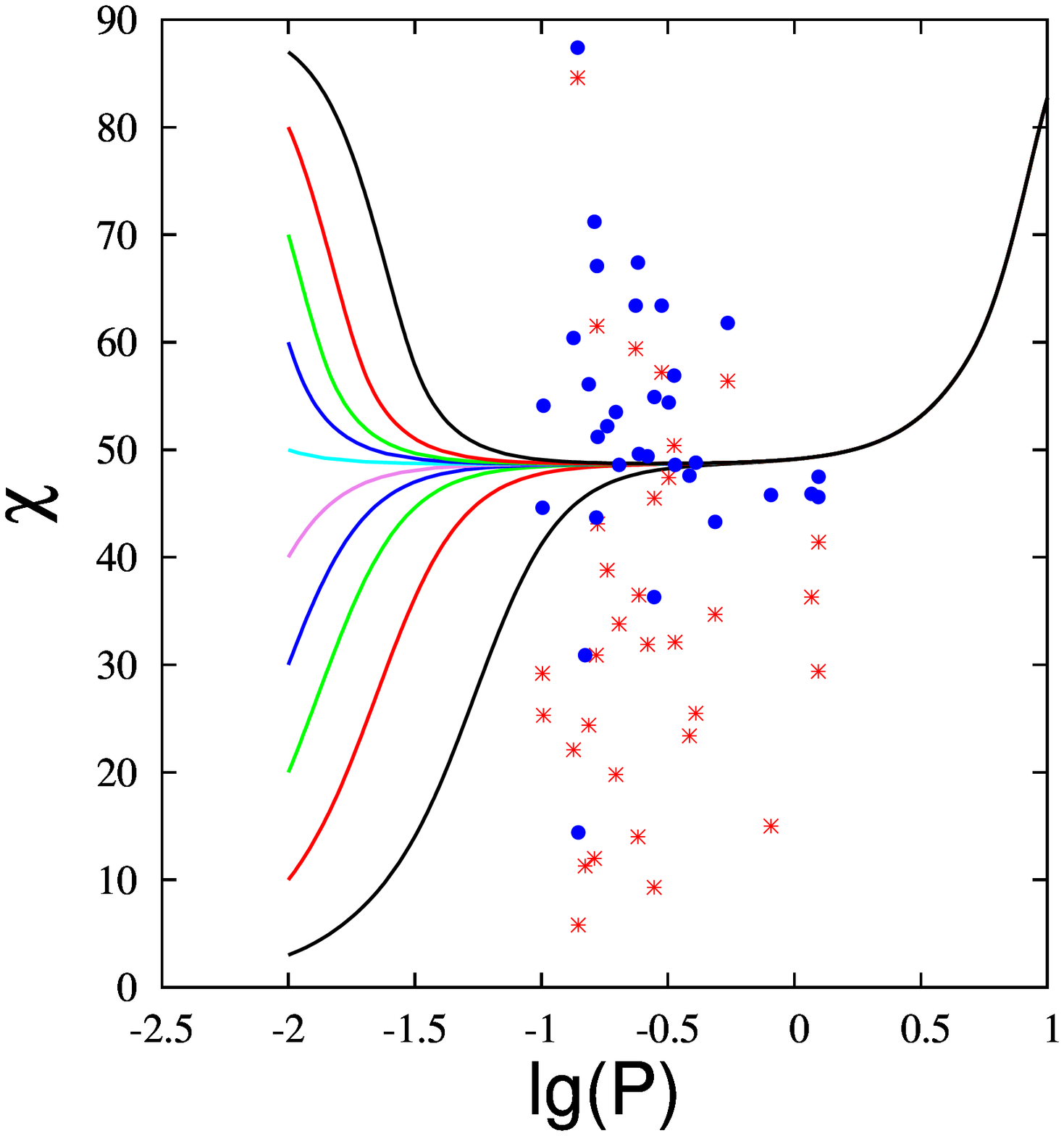}  
} % end parbox   
\caption{
   The same as in fig. \ref{fig_Icm2_init_1} but for
   $I_{c} / \tilde{I}_{tot} = 10^{-1}$,
   $I_{g} / \tilde{I}_{tot} = 10^{-5}$.
}
\label{fig_Icm1_init_1}
\end{figure}

\begin{figure}[h]
\parbox{0.45\hsize}{
 \includegraphics[angle=270,width=0.9\hsize]{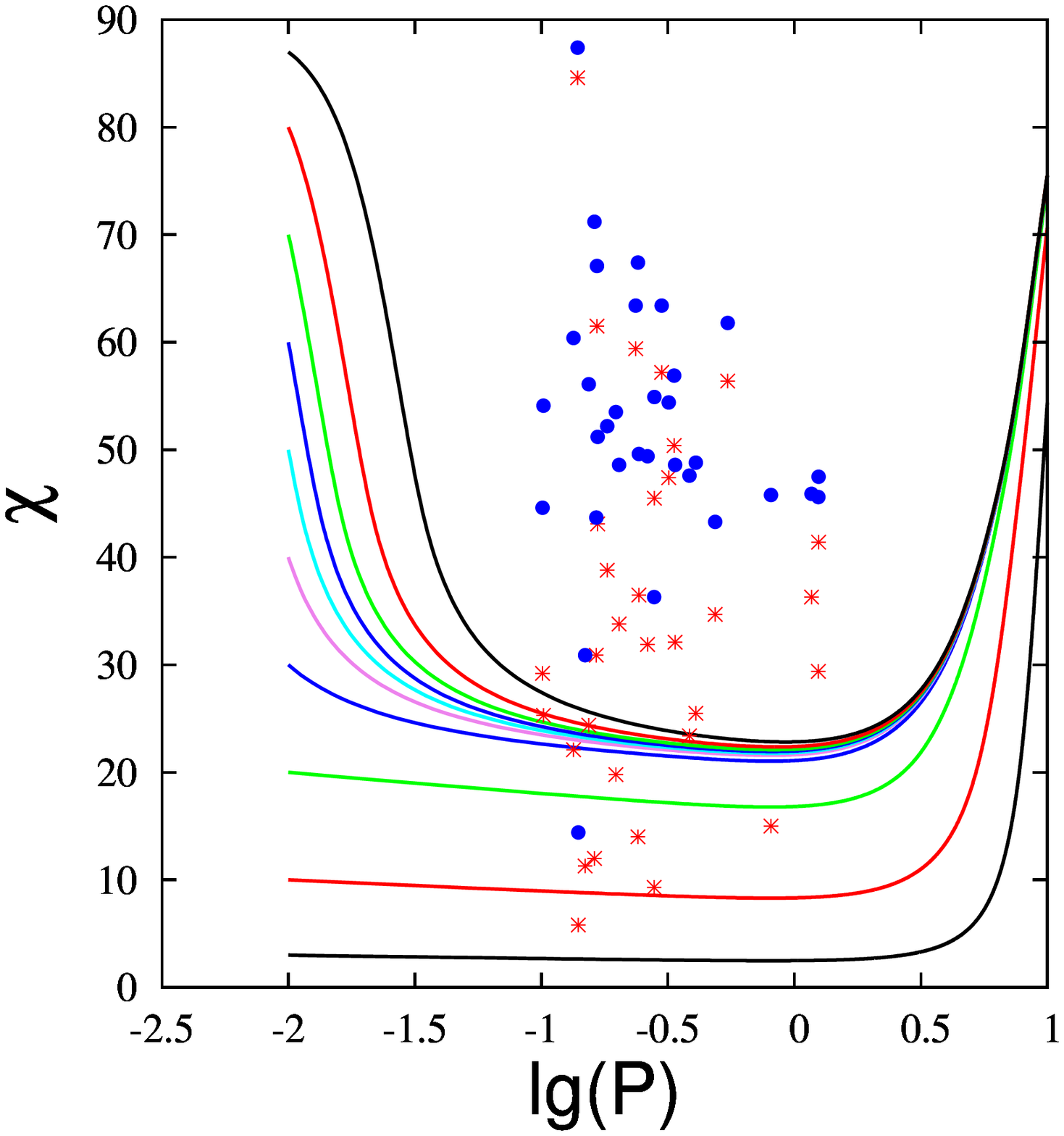} 
} % end parbox
\parbox{0.45\hsize}{
 \includegraphics[angle=270,width=0.9\hsize]{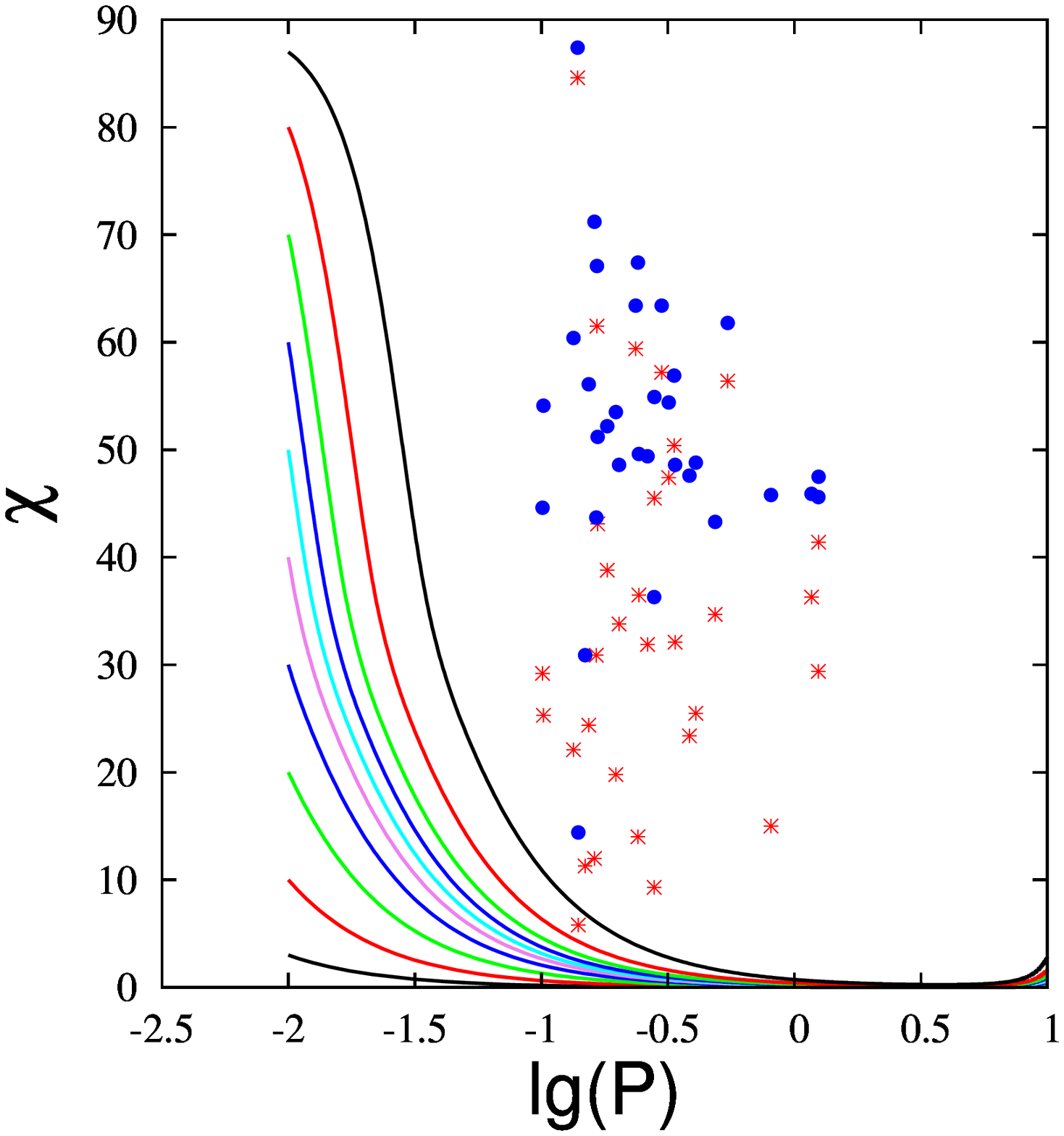}  
} % end parbox   
\caption{
   The same as in fig. \ref{fig_Icm2_init_2} but
   $I_{c} / \tilde{I}_{tot} = 10^{-1}$,
   $I_{g} / \tilde{I}_{tot} = 10^{-5}$.
   Left panel corresponds to $\nu = 0.8$,
   right panel to $\nu = 1.0$.    
}
\label{fig_Icm1_init_2}
\end{figure}

\begin{figure}[h]
\parbox{0.45\hsize}{
 \includegraphics[angle=270,width=0.9\hsize]{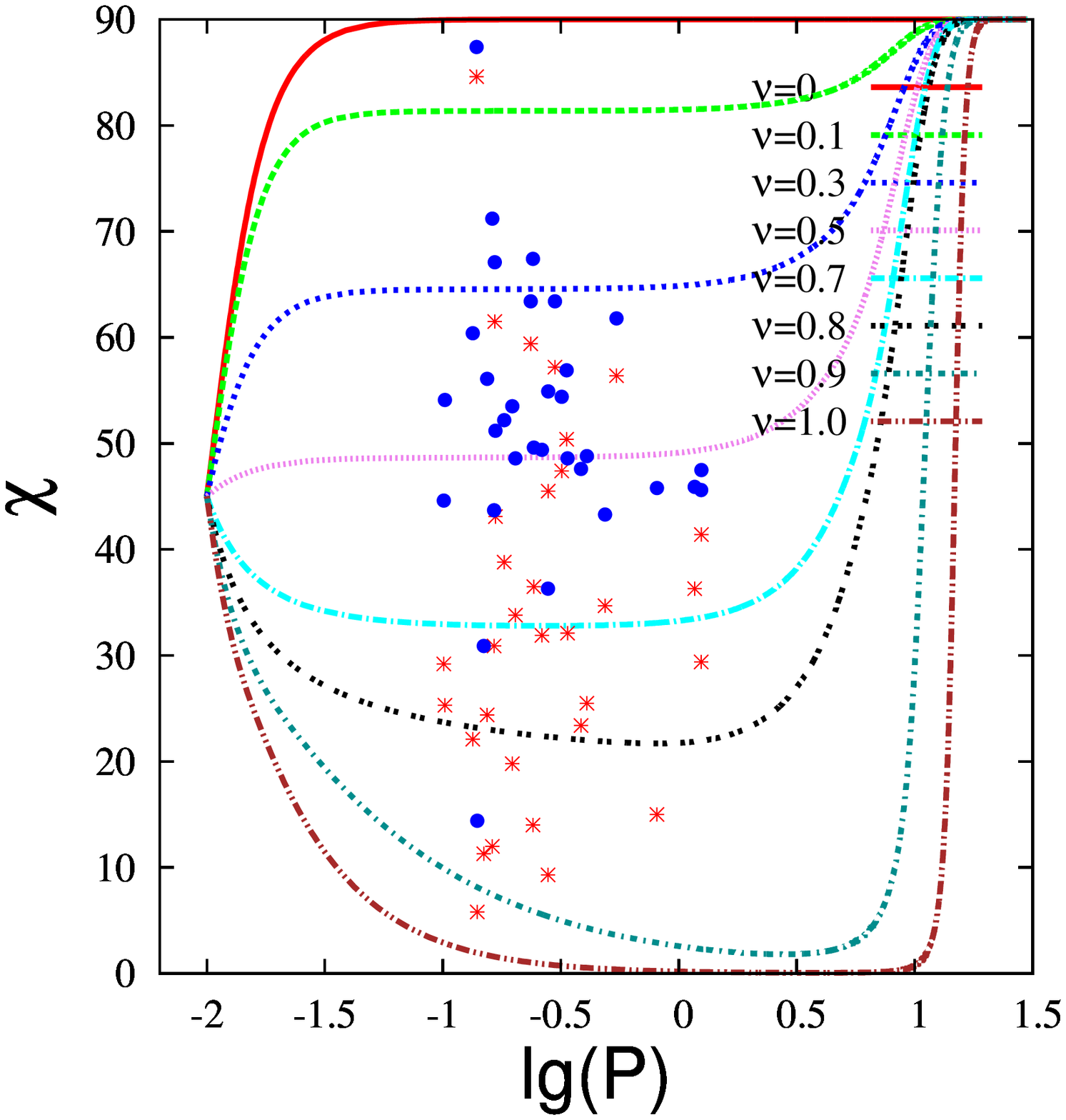} 
} % end parbox
\parbox{0.45\hsize}{
 \includegraphics[angle=270,width=0.9\hsize]{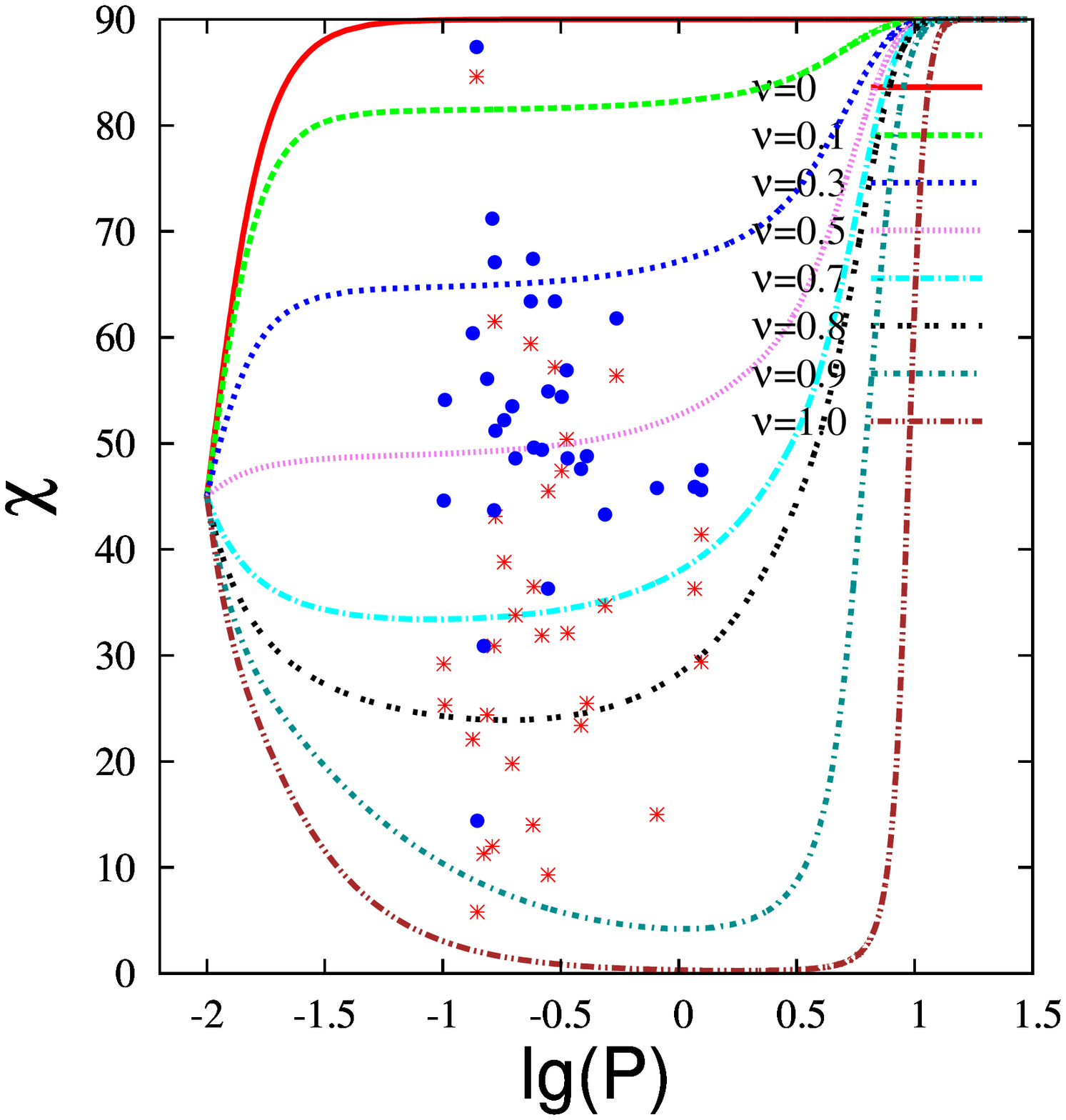}  
} % end parbox   
\caption{
   The same as in fig. \ref{fig_Icm2_nu} but
   $I_{c} / \tilde{I}_{tot} = 10^{-1}$,
   $I_{g} / \tilde{I}_{tot} = 10^{-5}$.
   Left panel corresponds to $\sigma = 10^{-10}$,
   right panel to $\sigma = 10^{-6}$. 
}
\label{fig_Icm1_nu}
\end{figure}

\begin{figure}[h]
\parbox{0.45\hsize}{
 \includegraphics[angle=270,width=0.9\hsize]{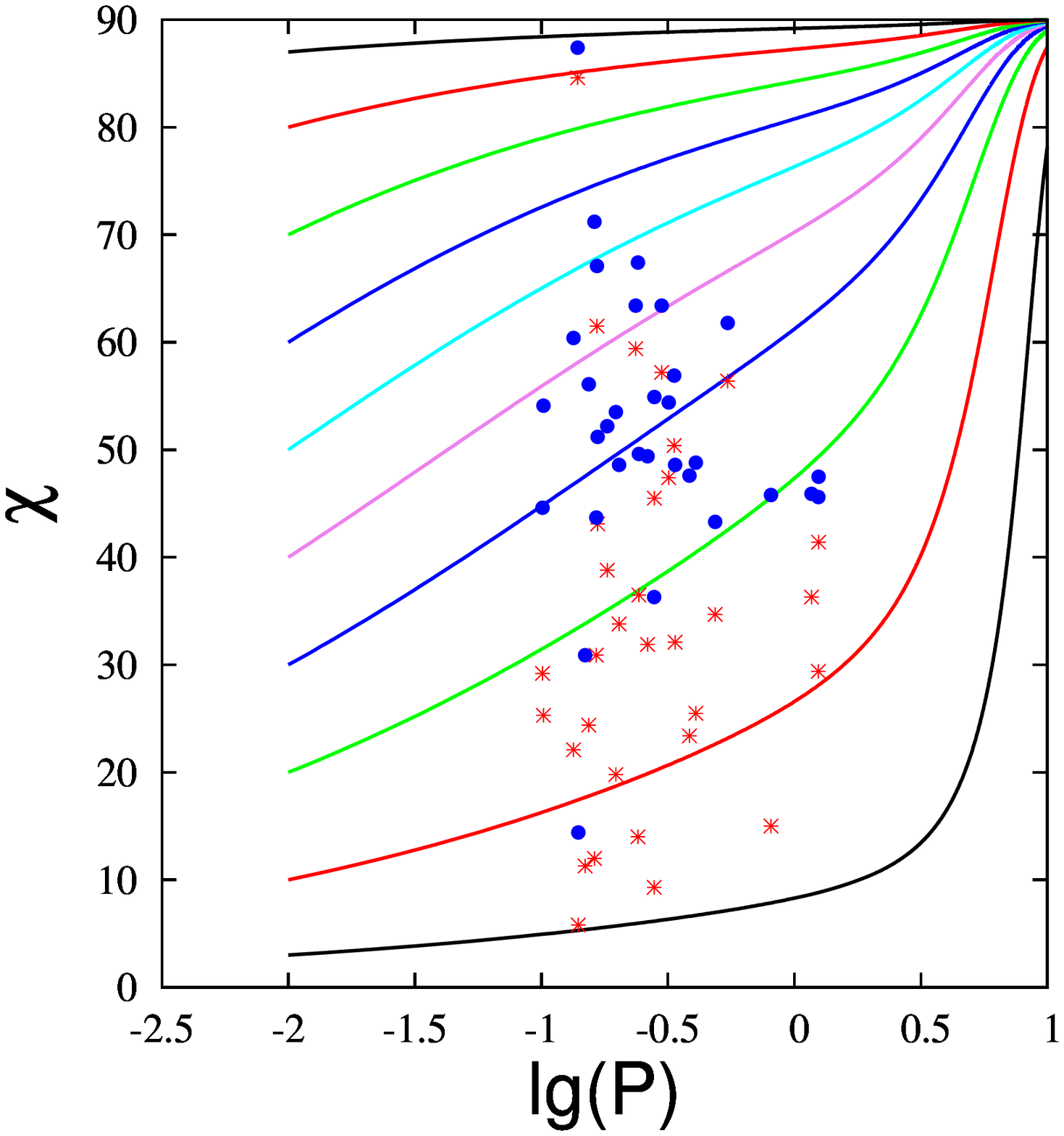} 
} % end parbox
\parbox{0.45\hsize}{
 \includegraphics[angle=270,width=0.9\hsize]{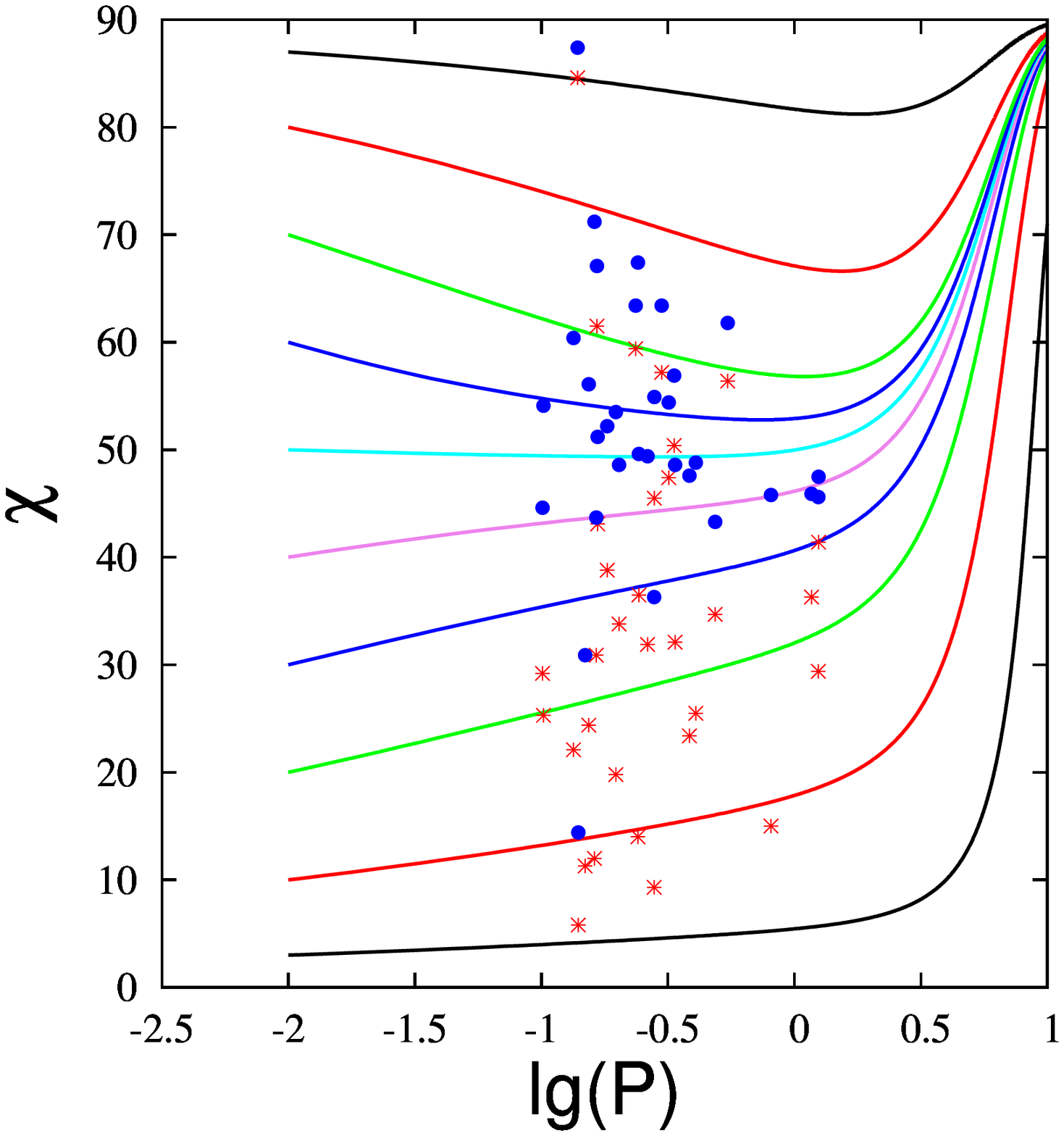}  
} % end parbox   
\caption{
   The same as in fig. \ref{fig_Icm2_init_1} but for
   $I_{g} / \tilde{I}_{tot} = 10^{-3}$,
   $I_{r} / \tilde{I}_{tot} = 10^{-3}$.
   Left panel corresponds to $\nu = 0$,
   right panel to $\nu = 0.5$.    
}
\label{fig_Icm0_init_1}
\end{figure}

\begin{figure}[h]
\parbox{0.45\hsize}{
 \includegraphics[angle=270,width=0.9\hsize]{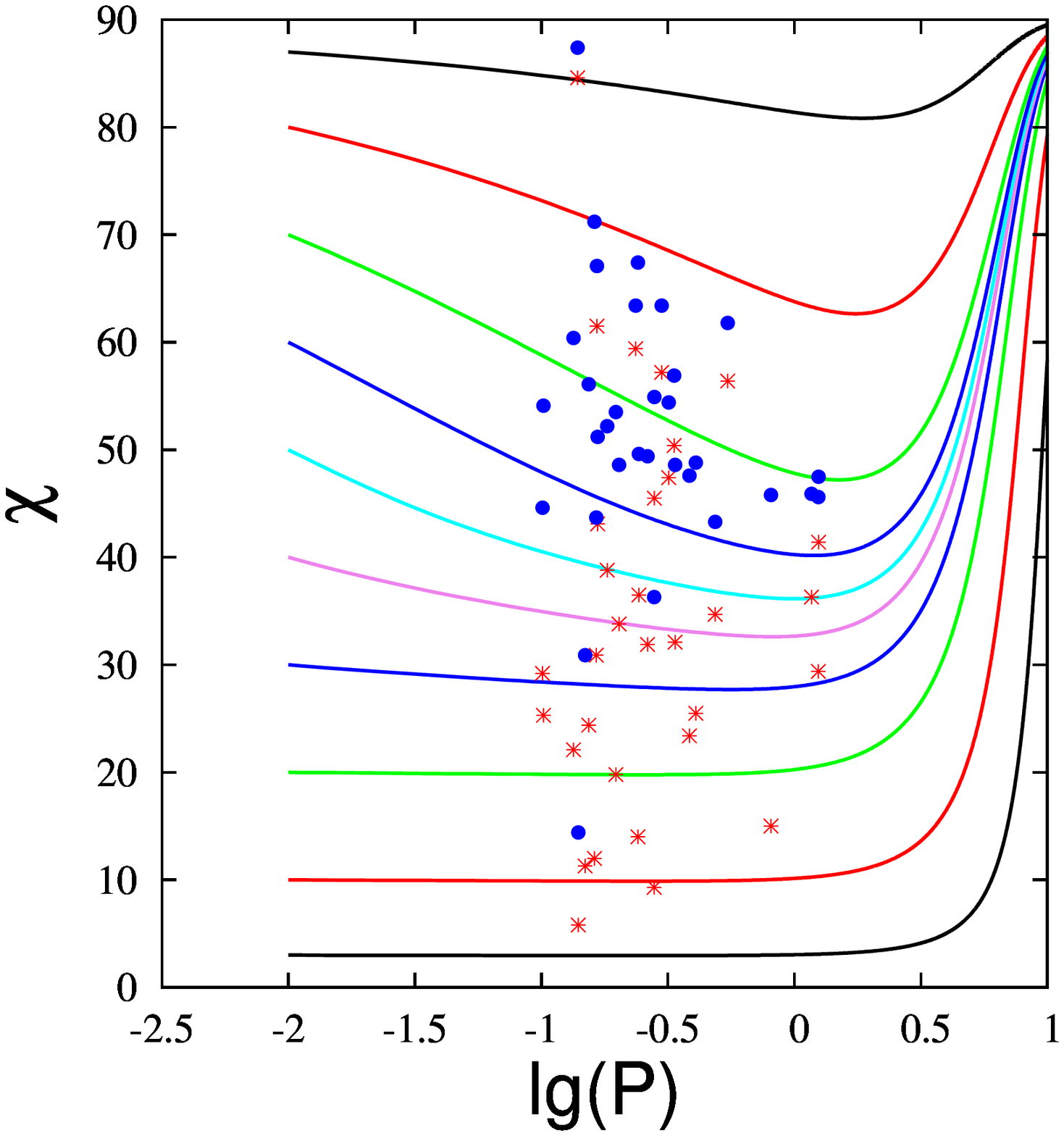} 
} % end parbox
\parbox{0.45\hsize}{
 \includegraphics[angle=270,width=0.9\hsize]{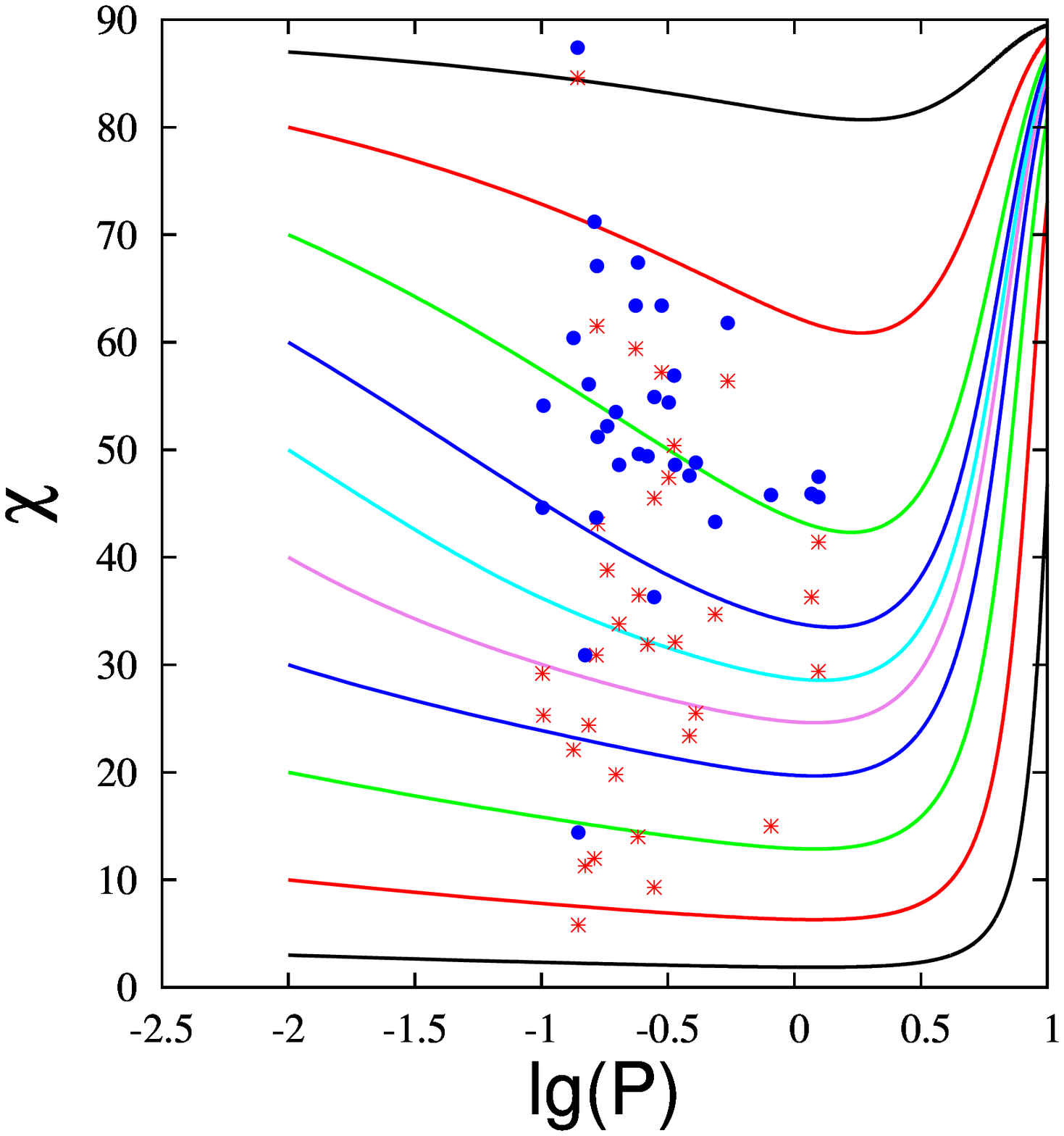}  
} % end parbox   
\caption{
   The same as in fig. \ref{fig_Icm2_init_2} but for
   $I_{g} / \tilde{I}_{tot} = 10^{-3}$,
   $I_{r} / \tilde{I}_{tot} = 10^{-3}$.
   Left panel corresponds to $\nu = 0.8$,
   right panel to $\nu = 1.0$.    
}
\label{fig_Icm0_init_2}
\end{figure}

\begin{figure}[h]
\parbox{0.45\hsize}{
 \includegraphics[angle=270,width=0.9\hsize]{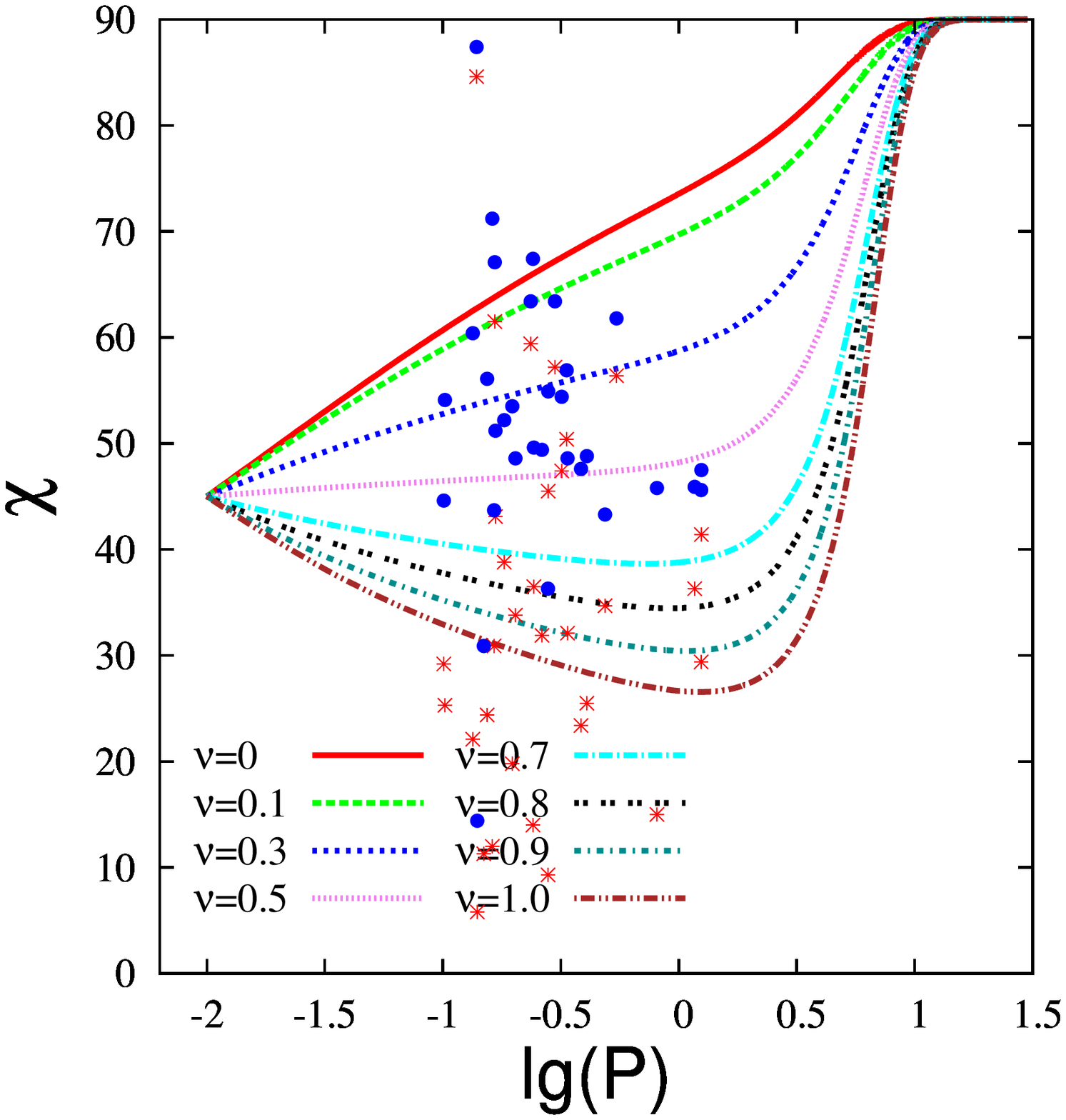} 
} % end parbox
\parbox{0.45\hsize}{
 \includegraphics[angle=270,width=0.9\hsize]{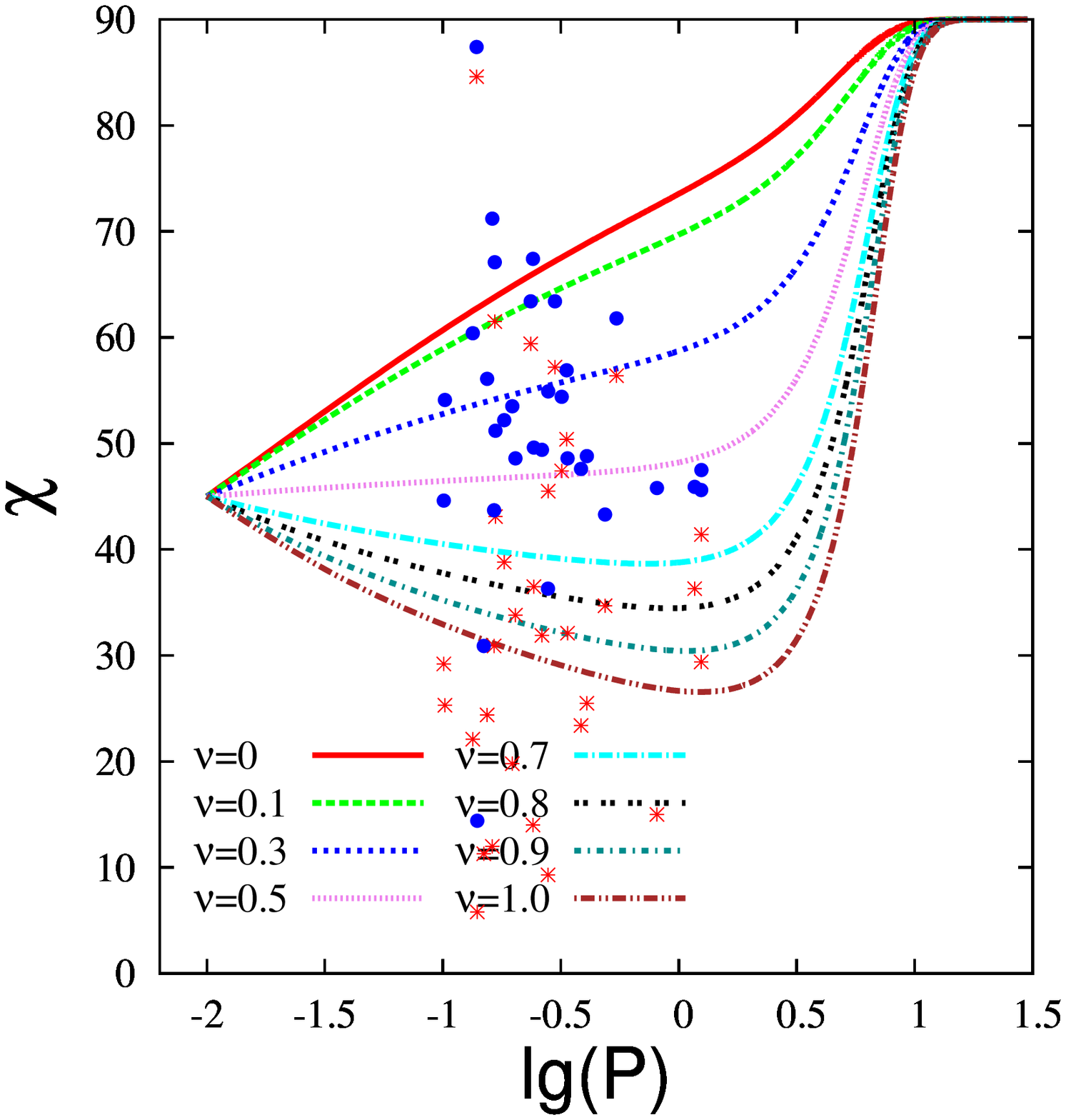}  
} % end parbox   
\caption{
   The same as in fig. \ref{fig_Icm2_nu} but for
   $I_{g} / \tilde{I}_{tot} = 10^{-3}$,
   $I_{r} / \tilde{I}_{tot} = 10^{-3}$.
   Left panel corresponds to $\sigma = 10^{-10}$,
   right panel to $\sigma = 10^{-6}$. 
}
\label{fig_Icm0_nu}
\end{figure}

In present paper we  will use the following component interaction model.
The crust and $g$-component interact with $r$-component
like normal matter with superfluid and
there are normal viscosity between the crust and $g$-component \cite{Sedrakian1999,Melatos2015} :
\begin{equation}
\beta_{cr} = \beta_{gr} = \Omega \frac{\sigma}{1+\sigma^{2}}
\mbox{, \ \ }
\alpha_{cr} = \alpha_{gr} = 2 \beta_{gc}
\mbox{, \ \ }
\gamma_{cr} = \gamma_{gr} = - \sigma \beta_{gc}
\mbox{, \ \ }
\beta_{cg} = \alpha_{cg}
\mbox{ \  and\ } 
\gamma_{cg} = 0.
\end{equation}
The solution of equation (\ref{PB_dchi_dP_base_eqn}) for
different initial inclination angles
$\chi$ and initial period $P = 10 \mbox{ ms}$
in the case of 
$\sigma = 10^{-10}$, $\alpha_{cg} = 10^{-1} s^{-1}$,
$I_{c} / \tilde{I}_{tot} = 10^{-2}$,
$I_{g} / \tilde{I}_{tot} = 10^{-3}$
is shown in fig. \ref{fig_Icm2_init_1} and \ref{fig_Icm2_init_2}.
We can see that in most cases 
the star forgets initial value of 
inclination angle $\chi$ very rapidly
and evolves to equilibrium inclination angle $\chi_{eq}$,
at which
\begin{equation}
\sin\chi \cos\chi \cdot
\left( B \, (1-< \alpha >) - \Gamma \, R_{eff} \right) \approx 0.
\end{equation}
The subsequent evolution of angle $\chi$ is caused by 
the slow changing of equilibrium angle $\chi_{eq}$ due to
pulsar braking and consequently growing of $R_{eff}$.
The solution (\ref{PB_dchi_dP_base_eqn}) for
initial inclination angle 
$\chi = 45^{\circ}$, initial period $P = 10 \mbox{ ms}$
and
different values of $\nu$,
in case of $\alpha_{cg} = 10^{-1} s^{-1}$,
$I_{c} / \tilde{I}_{tot} = 10^{-2}$,
$I_{g} / \tilde{I}_{tot} = 10^{-3}$
is shown in fig. \ref{fig_Icm2_nu}.
In this case $1-Q \approx 10^{-2}$.
The same but for
$I_{c} / \tilde{I}_{tot} = 10^{-1}$,
$I_{g} / \tilde{I}_{tot} = 10^{-5}$ 
is shown in fig \ref{fig_Icm1_init_1}-\ref{fig_Icm1_nu}.
In this case $Q \approx 0.9$. 
The increase of $I_{c}$ leads to slower evolution to
equilibrium angle $\chi_{eq}$.
The equilibrium inclination angle evolves slowly due to both
the decrease of $I_{g}$ and, hence, lesser dissipation, 
and the increase of $I_{c}$ and, hence,
large precession period $T_{p}$.
And, consequently, evolutionary tracks may pass
through the most pulsars.
The same but for
$I_{g} / \tilde{I}_{tot} = 10^{-3}$,
$I_{r} / \tilde{I}_{tot} = 10^{-3}$ 
is shown in fig. \ref{fig_Icm0_init_1} -  \ref{fig_Icm0_nu}. 
In this case, $I_{c} \approx I_{tot}$ and $Q \approx 10^{-3}$.
The star rotates almost as a rigid body.
Consequently, the inclination angle changes very slowly and
the pulsars usually do not reach the equilibrium angle
during their life.  


\section{Discussion}
We consider a model proposed in \cite{Sedrakian1999}
that allows simultaneously
the long-period precession and quasi-glitch events
with taking into account the influence of the small-scale
magnetic field on pulsar braking.
For simplicity we consider only the case of axial symmetric precession
and do not take into account that the presence of the small-scale magnetic
field makes precession triaxial
\cite{Goglichidze2015}. 
The main problem of proposed model is the exact nature of components,
especially, of $g$-component.
Strictly speaking, we only postulate
the existence of components with some properties.

\begin{figure}[h]
\parbox{0.45\hsize}{
    \includegraphics[width=0.9\hsize]{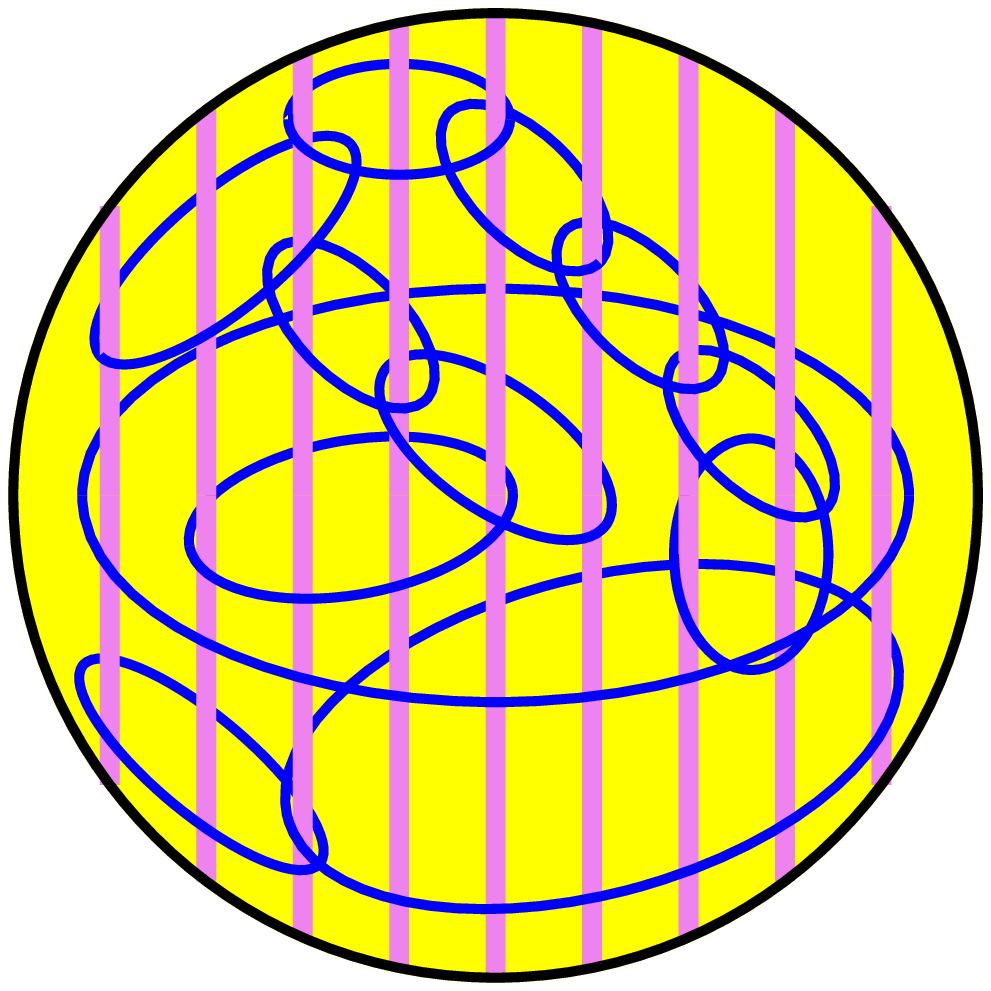}     
} % end parbox
\parbox{0.45\hsize}{
    \includegraphics[width=0.9\hsize]{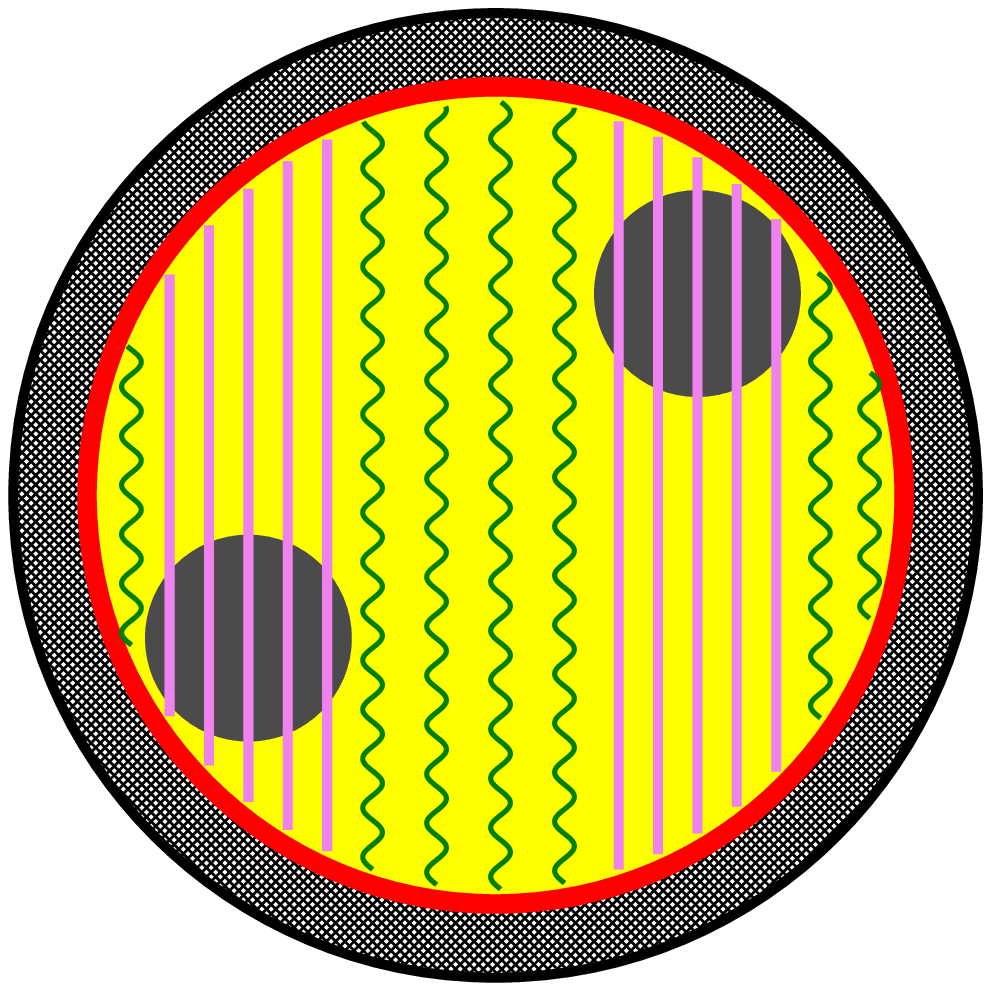}     
} % end parbox   
\caption{
The tangle of fluxoids with pinned vortices is show on the left.
Pinned vortices are shown by straight pink lines, closed fluxoids
are shown by blue lines.
Two  flowing in star core tangles are shown on the right.
Pinned vortices are shown by straight pink lines, 
free vortices are shown by wavy green lines.
The $g$-component is shown by dark gray circles and 
core superfluid outside $g$-component is shown by yellow area.
}
\label{fig_large_Ig}
\end{figure}

Let us speculate a little about 
possible nature of these components.
The $g$-component may consist of tangles of closed fluxoids,
normal matter inside the tangles and pinned superfluid 
(see fig. \ref{fig_large_Ig}).
These tangles "freely"\  flow inside the star core.
The collapse of closed fluxoids is prevented
by repulsion of pinned vortices.
Similar configurations are considered in 
\cite{Gurgercinoglu2014}, where vortices are pinned to fluxoids forming
the regular toroidal magnetic field inside the neutron star core,
and  \cite{Glampedakis2015}, where the "freely"\  flowing of magnetic field tangles is discussed.
We expect that $L_{g} \sim 0.1 \, I_{g}\Omega$.
Hence, if we suppose that $L_{g} \sim 10^{-2} I_{tot} \Omega$ \cite{Melatos2015}
then $I_{g} \sim 10^{-3} I_{tot}$.
In this case, $c$-component is exactly the cruct, so 
$I_{c} \sim 10^{-2} I_{tot}$ \cite{Yakovlev2007}
and $r$-component consists of normal and superfluid matter
located outside the tangles, so $I_{r} \approx I_{tot}$.
We suppose that strong interaction between the crust and the tangles
may be related to small number of fluxoids got out the tangles.
It is worth to note that a tangle located deep inside the core
weakly interacts with the crust and may produce
something like "slow glitch"\  \cite{Shabanova2007}.
The main problem of such configuration is the stability of tangles
and its partial destruction during glitches.

\begin{figure}[h]
\parbox{0.45\hsize}{
  \includegraphics[width=0.9\hsize]{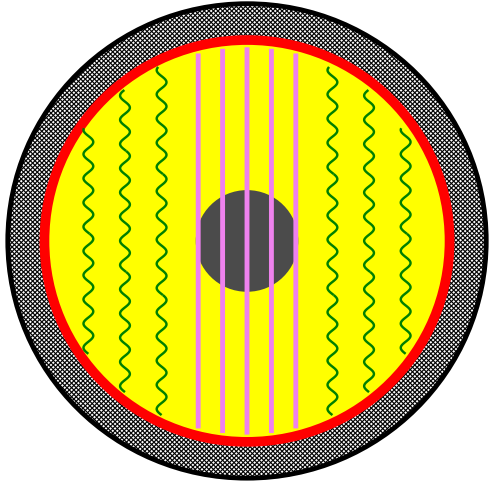}    
} % end parbox
\caption{
The star with small rigid core.
Vortices pinned to rigid core are shown by straight pink lines, 
free vortices are shown by wavy green lines.
The $g$-component is shown by dark gray circle and 
core superfluid outside the $g$-component is shown by yellow area. 
}
\label{fig_small_Ig}
\end{figure}

The $g$-component also may be created by small rigid core which can exist
in central region of the star \cite{Takatsuka1989} (see fig. \ref{fig_small_Ig}) with superfluid vortices pinned to it.
However, in this case, we must assume that the vortices are extremely rigid,
so a vortex pinned by its central part to the rigid core is fixed outside the rigid core as well.
Consequently, if we assume that the radius of rigid core
$r_{g} \sim (0.1-0.2) \, r_{ns}$ then
the moment of inertia of "normal"\  matter inside the rigid core
$I_{g} \sim 0.1 ( r_{g} / r_{ns} )^{3} \sim (10^{-6}-10^{-5}) \, I_{tot}$
However,  it controls the vortices movement and, hence, the superfluid flow 
in volume $ \sim r_{g}^{2} \cdot r_{ns}$, see fig. \ref{fig_small_Ig}.
Consequently, we can estimate the angular momentum of pinned superfluid:
$L_{g} \sim \left( r_{g} / r_{ns}  \right)^{2} I_{tot} \Omega
 \sim 10^{-2} I_{tot} \Omega$.
In this case, $c$-component consists of the crust
and normal matter of the core:
$I_{c} \sim 10^{-1} I_{tot}$ \cite{Yakovlev2007},
and $r$-component is superfluid which is not pinned 
to rigid core $I_{r} \approx I_{tot}$.
We suppose that the strong interaction 
between $g$-component and the crust
requires that the magnetic field penetrates from crust 
up to boundaries of rigid core.
Consequently, core superconductivity must be absent 
(at least outside the rigid core).

In the paper we assume that the interaction between
$c$-component and $g$-component is strong
in order to provide rapid angular momentum transfer
from $g$-component to the crust.
Hence, if $I_{g} \sim I_{tot}$, precession is damped
very rapidly and the inclination angle quickly evolves
to $\chi \approx 0^{\circ}$ or $\chi \approx 90^{\circ}$.
So in order to save precession during pulsar life time
it is necessary to suppose that $I_{g} \ll I_{tot}$.
We can reduce this restriction if we suppose
that the friction between the crust and $g$-component
increases during glitch.
For example, we can assume that the crust and $g$-component
interact due to weak viscous friction between two glitches
but during the the glitch the angular momentum is rapidly transfered 
by Kelvin waves \cite{Link2013},
Alfven or sound waves. 
We can also reduce the restriction if we assume
that $g$-component is slowly destroyed.
For example, the tangles of fluxoids slightly
destroyed during each glitch.

\ack
Authors thank O.A. Goglichidze and D.M.Sedrakian for
usefull discussion which has led to writting this paper,
A.N.Biryukov for usefull discussion of pulsar precession
and V.S.Beskin, I.F.Malov, E.B.Nikitina, A.N.Kazantsev, V.A. Urpin
and A.I.Chugunov for help and  useful discussions.


\end{document}